\documentclass[%
amsmath,amssymb,
aps,
]{revtex4-2}

\usepackage{graphicx}
\usepackage{epstopdf, epsfig}
\usepackage{graphicx}
\usepackage{dcolumn}
\usepackage{bm}
\usepackage{color}
\usepackage{array}
\usepackage{tabu}
\usepackage{amsmath,amssymb,stmaryrd}
\usepackage{subfigure}
\usepackage{subfigmat}
\usepackage{pbox}
\usepackage{xfrac}
\usepackage{tabularx}
\usepackage{multirow}
\usepackage[percent]{overpic}
\usepackage{float}

\usepackage[linesnumbered,ruled,vlined]{algorithm2e}
\usepackage{algorithmic}

\definecolor{blue2}{rgb}{0, 0.4470, 0.7410}
\definecolor{red2}{rgb}{0.8500, 0.1250, 0.0480} 
\definecolor{orange}{rgb}{0.8500, 0.3250, 0.0980} 
\definecolor{yellow}{rgb}{0.9290, 0.6940, 0.1250}
\definecolor{purple}{rgb}{0.4940, 0.1840, 0.5560}
\definecolor{green}{rgb}{0.4660, 0.6740, 0.1880}
\definecolor{ltblue}{rgb}{0.3010, 0.7450, 0.9330}
\definecolor{dkred}{rgb}{0.6350, 0.0780, 0.1840}
\definecolor{gray}{rgb}{0.22, 0.22, 0.3}

\newcolumntype{M}[1]{>{\centering\arraybackslash}m{#1}}

\makeatletter
\def\BState{\State\hskip-\ALG@thistlm}
\makeatother

\def\drawline#1#2{\raise 2.0pt\vbox{\hrule width #1pt height #2pt}}

\newcommand{\bs}{\boldsymbol}

\usepackage{bm}

\begin{document}


\title{Randomized resolvent analysis}

\author{Jean H\'elder Marques Ribeiro}
 \email{jeanmarques@ucla.edu}
 \affiliation{Department of Mechanical and Aerospace Engineering, 
 University of California, Los Angeles, CA 90095, USA}
 
\author{Chi-An Yeh}
 \email{cayeh@seas.ucla.edu}
 \affiliation{Department of Mechanical and Aerospace Engineering, 
 University of California, Los Angeles, CA 90095, USA}

\author{Kunihiko Taira}
 \email{ktaira@seas.ucla.edu}
 \affiliation{Department of Mechanical and Aerospace Engineering, 
 University of California, Los Angeles, CA 90095, USA}

\date{\today}

\begin{abstract}
Performing global resolvent analysis for high-Reynolds-number turbulent flow calls for the handling of a large discrete operator.  
Even though such large operator is required in the analysis, most applications of resolvent analysis extracts only a few dominant resolvent response and forcing modes.
Here, we consider the use of randomized numerical linear algebra to reduce the dimension of the resolvent operator for achieving computational speed up and memory saving compared to the standard resolvent analysis.  To accomplish this goal, we utilize sketching of the linear operator with random test matrices with a Gaussian distribution and with insights from the base flow incorporated to perform singular value decomposition on a low-rank matrix holding dominant characteristics of the full resolvent operator.  The strength of the randomized resolvent analysis is demonstrated on a turbulent separated flow over an airfoil.  This randomized approach clears the path towards tackling resolvent analysis for higher-Reynolds number bi- and tri-global base flows.

\begin{description}
\item[Keywords]
resolvent analysis,
instabilities,
randomized methods
\end{description}

\end{abstract}


\maketitle

\section{\label{sec:intro}Introduction}

One of the central questions in turbulence and flow control is concerned with the evolution of perturbations.  Gaining detailed understanding of the perturbation dynamics in turbulent flows is a daunting task, due to the complex nonlinear dynamics that takes place over a broad range of spatial and temporal scales.  To modify the global flow characteristics with control, it is necessary that the nonlinear interactions involving the actuation input (perturbation) become appropriately large to alter the base flow.  For flow control, we need not track all possible ways in which the actuation input can modify the flow, but instead we can focus on the dominant directions in which the perturbation can be amplified.  This notion has led to the modal analysis based approaches \citep{Holmes12, Schmid:JFM10, Taira:AIAAJ17,Taira:AIAAJ19}, including the global stability analysis \citep{Theofilis:ARFM11} and the resolvent analysis \citep{Trefethen:Science93}.  

In the presence of sustained perturbations or forcing inputs, the linear system response can be described by the transfer function from control theory.  This linear analysis is greatly simplified when the input to the system is sinusoidal and leads to the well-known Bode plots that reveal the gain and phase response of the system over a range of forcing frequencies.  The transfer function that relates the system input to the output is called the resolvent and its analysis has been extended to fluid flows by \citet{Trefethen:Science93}.  The resolvent analysis is based on the pseudospectral analysis and has been used to study the transient energy growth \citep{Trefethen:Science93} as well as the harmonic response of the system \citep{Jovanovic:JFM05}.  These initial studies of resolvent analysis were performed about stable laminar flows.  

An extension to resolvent analysis to study turbulent flows was presented by \citet{McKeon:JFM10}.  They considered the nonlinear advection term to be the self-sustained input within the natural feedback loop of the fluid flow.  This viewpoint has enabled the use of time-averaged base flows to reveal the input-output dynamics of turbulent flows.  Moreover, discounting or finite-time horizon based extension of the resolvent analysis has enabled resolvent analysis to study flows with unstable base states \citep{Jovanovic:Thesis2004, Yeh:JFM19}.  Because resolvent analysis can determine the most amplified forcing and response directions, it serves as a powerful analytical tool to find effective active and passive flow control techniques \citep{Yeh:JFM19, Nakashima:JFM17}.  

The resolvent analysis needs two key ingredients: (i) the base flow and (ii) the linearized Navier--Stokes operator.  It is known that the accuracy of the base flow and the spatial discretization of the linear operators is critical for extracting response characteristics correctly \citep{Yeh:JFM19}.  The need for accurate discretization of the linearized Navier--Stokes operators calls for sufficient grid resolution and appropriate computational domain size.  As such, the discrete resolvent operator becomes large with size $m \times m$. Here, $m$ is essentially the number of variables times the size of the grid, which can easily be upward of $\mathcal{O}(10^6)$ for turbulent flows \citep{Kajishima17}.  For resolvent analysis, the singular value decomposition (SVD) needs to be performed on the large resolvent operator with a taxing operation count of $\mathcal{O}(m^3)$. To enable resolvent analysis at high Reynolds numbers, we must find a relief to perform SVD of the resolvent operator.

Although the resolvent analysis is performed on a very large matrix, only the leading forcing and response modes are generally sought.  Based on the amount of necessary matrix data used to perform the analysis, the desired output is only a very small fraction of the input data size.  For this reason, it would be natural to consider that all elements of the resolvent matrix are not necessary to determine the leading resolvent modes.  Within the resolvent framework, the core of the computations lies with the SVD.  In order to handle a large operator for SVD to find the leading modes, we can consider subsampling the matrix of interest and perform the SVD on the low-order representation of this large matrix.  Randomized numerical linear algebra has recently emerged as an effective technique to reduce a large matrix to its low-order representation \citep{Halko:SIAMReview11, Drineas:ACM16, Tropp:SIAMMAA17}, for applications including big data compression and data transfer. The key idea is to pass a randomly generated low-rank test matrix through the large matrix to obtain the so-called sketch of the full matrix.  This sketch is low-rank but holds key information about the full matrix and can be used to derive appropriate bases to represent the full matrix in a low-dimensional manner \citep{Clarkson:ACM07, Woolfe:ACHA08, Halko:SIAMReview11, Tropp:SIAMMAA17}.  Randomized techniques can be incorporated into SVD to achieve tremendous computational and memory savings \citep{Woolfe:ACHA08, Rocklin:SIAMJMAA09, Halko:SIAMReview11}.

In recent years, modal analyses are tackling flows over complex geometries and high-Reynolds number flows with increasingly large degrees of freedom \citep{Taira:AIAAJ17,Taira:AIAAJ19,Jeun:POF16,Schmidt:JFM18,Dwivedi:JFM19,Ribeiro:POF17,Ricciardi:18AIAAJ}.  To further aid this endeavor, randomized SVD have been incorporated into data-based modal analysis techniques, including the proper orthogonal decomposition \citep{Rocklin:SIAMJMAA09} and dynamic mode decomposition \citep{Erichson:arXiv17}.  For global operator-based analyses of high-Reynolds-number flows \citep{Jeun:POF16,Schmidt:JFM18,Dwivedi:JFM19}, the leading singular values and modes can be determined with significant reduction in computational costs with the aid of randomized techniques.  The randomized technique presented in this study can greatly expand the applicability of the resolvent analysis to high-Reynolds number flows with multiple inhomogeneous spatial directions.  

In fact, randomized SVD has been adopted in the one-dimensional resolvent analysis of turbulent channel flow by \citet{Moarref:JFM2013}.  Since the details on the use of the randomized techniques for large-scale resolvent analysis is not available, one of our objectives is to provide detailed guidance on the use of randomized resolvent analysis for multi-dimensional turbulent base flows.  Moreover, the error bound of randomized SVD is generally not guaranteed and needs to be further characterized.  We also note that the sketching strategy can be further improved with insights from the base flow.  In the present study, we aim to address these issues on the use of randomized SVD in the context of resolvent analysis for fluid flows.  

In what follows, we present the randomized resolvent analysis and demonstrate its use for turbulent flow over a canonical airfoil.  In section \ref{sec:approach}, we introduce the randomized resolvent analyses and discuss how randomized SVD can be implemented without the explicit generation of inverse matrices.  We also propose an alternative route of recovering the response modes (left singular vectors) with significantly enhanced accuracy.  In section \ref{sec:example}, the present randomized technique is applied to the resolvent analysis of turbulent flow over a NACA 0012 airfoil.  We assess the level of error and the convergence behavior with respect to resolvent modes and gains obtained from the use of randomized SVD.  To further enhance the accuracy of randomized resolvent analysis, we introduce a physics-informed sampling technique that leverages the insights from the base flow.  At last, we provide concluding remarks in section \ref{sec:conclusion}.


\section{Approach}
\label{sec:approach}

\subsection{Full resolvent analysis}
\label{sec:resolvent}

Let us consider the flow state $\pmb{q} \in \mathbb{R}^{m}$ as a sum of the time-invariant base state $\bar{\pmb{q}}$ and the statistically stationary fluctuating component $\pmb{q}'$.  With this Reynolds decomposed flow variable and appropriate discretization, we can express the discrete Navier--Stokes equations as
\begin{equation}
    \frac{\partial \pmb{q}'}{\partial t} = \bm{L}_{\bar{\pmb{q}}} \pmb{q}'+ \pmb{f}',
    \label{eq:re-dec-final}
\end{equation}
where $\bm{L}_{\bar{\pmb{q}}} \in \mathbb{R}^{m\times m}$ is the linearized Navier--Stokes operator about the base state $\bar{\pmb{q}}$ and $\pmb{f}'$ collects the nonlinear terms and the external forcing inputs.  We gather the nonlinear terms as external forcing in the turbulent mean flow following the perspective of \citet{McKeon:JFM10}, \citet{Farrel:PRL94}, and \citet{Schmid:AR07}.  For the traditional resolvent analysis, $\bar{\pmb{q}}$ is chosen to be the stable laminar equilibrium state such that $\pmb{f}'$ can be considered as the forcing input to the system with the nonlinear term neglected \citep{Jovanovic:JFM05}.  More recently, turbulent mean flows has been used for ${\bar{\pmb{q}}}$ with $\pmb{f}'$ representing the nonlinear terms as sustained forcing input within the natural feedback system \citep{McKeon:JFM10}.

We can consider the Fourier transform $\left[\pmb{q}'(\pmb{x}, t),  \pmb{f}'(\pmb{x}, t)\right] = 
	\int_{-\infty}^{\infty}
	    [\hat{\pmb{q}}_{\omega}(\pmb{x}), \hat{\pmb{f}}_{\omega}(\pmb{x})] 
    e^{- i\omega t} {\rm d}\omega$ and express the relationship between $\pmb{q}'$ and $\pmb{f}'$ in frequency space as
\begin{equation}
    -i\omega \hat{\pmb{q}}_\omega = \bm{L}_{\bar{\pmb{q}}} \hat{\pmb{q}}_\omega + \hat{\pmb{f}}_\omega,
   \label{eq:re-dec-fourier}
\end{equation}
where $\omega$ is the frequency.  Note that spatial Fourier transform can also be incorporated if directional homogeneity is present.  For stable base flows, $\omega$ can be chosen to be real.  To extend resolvent analysis to unstable base flows, we can consider the use of finite-time/discounted analysis \citep{Jovanovic:Thesis2004, Yeh:JFM19} by choosing a complex frequency $\omega = \omega_r + i\beta$, where both $\omega_r$ and $\beta$ are real and $\beta$ discounts the modal growth rate of $\bm{L}_{\bar{\pmb{q}}}$.  The input-output relationship between $\hat{\pmb{f}}$ and $\hat{\pmb{q}}$ can be found from (\ref{eq:re-dec-fourier}) as
\begin{equation}
	\hat{\pmb{q}}_\omega 
	= \bm{A}
	\hat{\pmb{f}}_\omega,
	\label{eq:ns-resol}
\end{equation}
where 
\begin{equation}
    \bm{A} = \left[ -i\omega\bm{I} - \bm{L}_{\bar{\pmb{q}}} \right]^{-1} 
    \in \mathbb{C}^{m \times m}
    \label{eq:resol-op}
\end{equation}
is referred to as the {\it resolvent operator} \citep{Trefethen:Science93,Jovanovic:JFM05,McKeon:JFM10}.  It serves as a transfer function that amplifies (or attenuates) the harmonic forcing input $\hat{\pmb{f}}_\omega$ and maps it to the response $\hat{\pmb{q}}_\omega$.  The goal of resolvent analysis is to identify the dominant directions along which $\hat{\pmb{f}}_\omega$ can be most amplified through $\bm{A}$ to form the corresponding responses in $\hat{\pmb{q}}_\omega$.  This question is addressed by the SVD of 
\begin{equation}
    \bm{A} = \bm{U}\pmb{\Sigma}\bm{V}^*,
    \label{eq:svd}
\end{equation}
where $\bm{V}^*$ denotes the Hermitian of $\bm{V}$.  Resolvent analysis interprets left and right singular vectors $\bm{U} = [\hat{\pmb{u}}_1, \hat{\pmb{u}}_2, \dots, \hat{\pmb{u}}_m] \in \mathbb{C}^{m \times m}$ and $\bm{V} = [\hat{\pmb{v}}_1, \hat{\pmb{v}}_2, \dots, \hat{\pmb{v}}_m]  \in \mathbb{C}^{m \times m}$ respectively as response modes and forcing modes, with the magnitude-ranked singular values $\pmb{\Sigma} = {\sf{diag}}(\sigma_1, \sigma_2, \dots, \sigma_m) \in \mathbb{R}^{m \times m}$ being the amplification (gain) for the corresponding forcing-response pair.  For unstable base flows, it is important that a finite-time window is chosen with $\beta$ larger than the highest growth rate such that the resolvent analysis reveals the input-output relationship on a shorter time scale than that of the base flow instability \citep{Jovanovic:Thesis2004, Yeh:JFM19}.

Performing the SVD of $\bm{A} \in \mathbb{C}^{m\times m}$ requires an theoretical operation count of $\mathcal{O}(m^3)$. In practice, some algorithms can reduce this operation count when only a few singular values are to be recovered, while still being computationally taxing for large $m$ \citep{Morgan:LinAlg2006}.  Such cases are encountered in high-Reynolds number flows and bi/tri-global analysis settings.  However, we note that many applications of resolvent analysis call only for the dominant forcing and response modes $[\hat{\pmb{v}}_1, \hat{\pmb{u}}_1]$ associated with the highest gain $\sigma_1$.  This is appropriate when the first gain $\sigma_1$ is much larger than the rest of the gains $\sigma_{j>1}$ and shows a quick roll off. Such condition is related to non-normality of linear operator $\bm{L}_{\bar{\pmb{q}}}$, which is encountered for flows with strong shear and separation \cite{Schmid01}.

When the contributions from the higher-order modes are neglected, it is referred to as the rank-1 assumption for which the flow response from forcing $\hat{\pmb{f}}$ is approximated as $\hat{\pmb{q}} \approx \hat{\pmb{u}}_1 \sigma_1 \langle \hat{\pmb{v}}_1, \hat{\pmb{f}} \rangle$, provided that $\sigma_1 \gg \sigma_2$ and $\hat{\pmb{f}}$ has reasonable magnitude along $\hat{\pmb{v}}_1$.  For seeking only the dominant modal insights from resolvent analysis, we discuss a remedy for performing large-scale resolvent analysis \citep{Jeun:POF16, Schmidt:JFM18} in a computationally tractable manner below.


\subsection{Randomized resolvent analysis}
\label{sec:random}

For a flow that have dominant structures, we consider a low-rank representation of the resolvent operator $\bm{A}$.
That is, instead of directly performing SVD of $\bm{A}$ and obtaining the leading-mode representation, we seek a low-rank representation of $\bm{A}$ and perform the SVD of the low-rank version of $\bm{A}$.
We can consider finding an appropriate low-dimensional basis to project the large resolvent operator on a suitable subspace to derive the low-rank resolvent approximation.  

The action of a full matrix on a vector should reveal some insights on which components are modified in the dominant directions.  In the case of flow that can be described with the rank-1 approximation, there should be a low number of dominant directions.  This very point can be taken advantage of through what is known as {\it sketching} in numerical linear algebra.  Sketching refers to a procedure in which a tall and skinny test matrix $\bs{\Omega} \in \mathbb{R}^{m \times k}$ (or $\mathbb{C}^{m \times k}$), where $k \ll m$, is passed through $\bm{A}$
\begin{equation}
    \bm{Y} = \bm{A}\bs{\Omega}.
\end{equation}
Here, matrix $\bm{Y} \in \mathbb{C}^{m \times k}$ is called the {\it sketch} of the input matrix $\bm{A}$ \citep{Woolfe:ACHA08, Halko:SIAMReview11, Tropp:SIAMMAA17}.  
The test matrix $\bs{\Omega}$ can be constructed using random values with Gaussian distribution \citep{Martinsson:ACHA2011} and be weighted by any input matrix insight, as will be discussed in section \ref{sec:physicsomega} for a physics-inspired random test matrix.  
As the sketch holds the dominant influence of $\bm{A}$, we can consider orthonormalizing $\bm{Y}$ using a QR decomposition to form the orthonormal basis with $\bm{Q} \in \mathbb{C}^{m \times k}$ upon which we can project the full matrix $\bm{A}$ to derive its low-rank approximation.  
In this way, it is possible to approximate $\bm{A}$ for a rank $k \ll m$ as long as this approximation preserves the features of the leading modes.

Given this $\bm{Q}$, a low-rank approximation of $\bm{A}$ can be found as $\bm{A} \approx \bm{Q}\bm{Q^*} \bm{A}$ \citep{Halko:SIAMReview11}.  We can view this as a low-rank decomposition of $\bm{A} \approx \bm{Q} \bm{B}$, where $\bm{B} = \bm{Q^*} \bm{A} \in \mathbb{C}^{k\times m}$.  It is this reduced matrix $\bm{B}$ upon which we can perform the SVD
\begin{equation}
	\boldmath{B} = \bm{\tilde{U}}\boldsymbol{\Sigma}\boldmath{V^*}
\end{equation}
Hence, as a low-rank approximation, we now have 
\begin{equation} 
	\bm{A}
	\approx \bm{Q}\bm{\tilde{U}}\boldsymbol{\Sigma}\bm{V^*}
	\label{eq:approx}
\end{equation}
where we can consider $\bm{U} \approx \bm{Q}\bm{\tilde{U}}$.  This process is the {\it randomized SVD} \citep{Halko:SIAMReview11}, where the sketch $\bm{Y}$ was used to derive $\bm{Q}$.  This approximation almost always satisfies
$
	\| \bm{A} - \bm{Q}\bm{Q^*} \bm{A}\|
	\le \left( 1+ 9\sqrt{k+p} \sqrt{m} \right) \sigma_{k+1}
$,
where $p$ is the oversampling parameter, which is applied to build an approximation of rank $k$ while projecting the matrix $\bm{A}$ to the low-dimensional subspace with $(k+p)$ vectors.
With this overall approach, the computational cost for SVD is reduced to $\mathcal{O}(m k^2)$ instead of $\mathcal{O}(m^3)$ for the full SVD. 
In our implementation, we use $\bs{v}_1$ from (\ref{eq:approx}) and retrieve the leading singular value and left singular vector through 
\begin{equation}
    \bm{A} \hat{\bs{v}}_1 = \hat{\bs{u}}_1 \sigma_1.
	\label{eq:U=AVS}
\end{equation}
The singular value and vector can be separated by noticing that $\| \hat{\bs{u}}_1 \| = 1$.  Applied to resolvent analysis, the last equation provides more accurate leading singular value $\sigma_1$ and left singular vector $\hat{\bs{u}}_1$ compared to the original randomized SVD algorithm \citep{Halko:SIAMReview11}. The same operation can be used to recover the higher-order modes, with better accuracy than using the original algorithm \citep{Halko:SIAMReview11}. For applications where high-order modes and orthogonality are desired, we can solve for $\bm{U}\bs{\Sigma}$ and compute its SVD. In the present randomized resolvent analysis, we emphasize that only the discrete linear operator $\bm{L}_{\bar{\pmb{q}}}$ is needed for sketching $\bm{Y}$ and to find the reduced matrix $\bm{B}$.  Unlike the original resolvent method, matrix linear solvers are used to avoid calling for the inverse within the resolvent operator.  The resulting algorithm constitutes the {\it randomized resolvent analysis} summarized in Algorithm 1.

To utilize the randomized SVD for resolvent analysis, we must be aware that the resolvent operator $\bm{A} = \left[ -i\omega\bm{I} - \bm{L}_{\bar{\pmb{q}}} \right]^{-1}$ contains an inverse operation in its definition, which need not be numerically performed. We do not intend to perform an inverse operation within $\bm{A}$ in the present work. In the full resolvent analysis, when the matrices become too large and the inverse can not be performed (which is likely the case for 2D and 3D problems), one can focus on modes corresponding to the smallest singular values of $\bm{A}^{-1}$ to find those for the largest singular values of $\bm{A}$. Similar approaches have avoided the inverse computation, including the work by \citet{Jeun:POF16}. 

\begin{figure}
    \centering
    \includegraphics[width=1\textwidth]{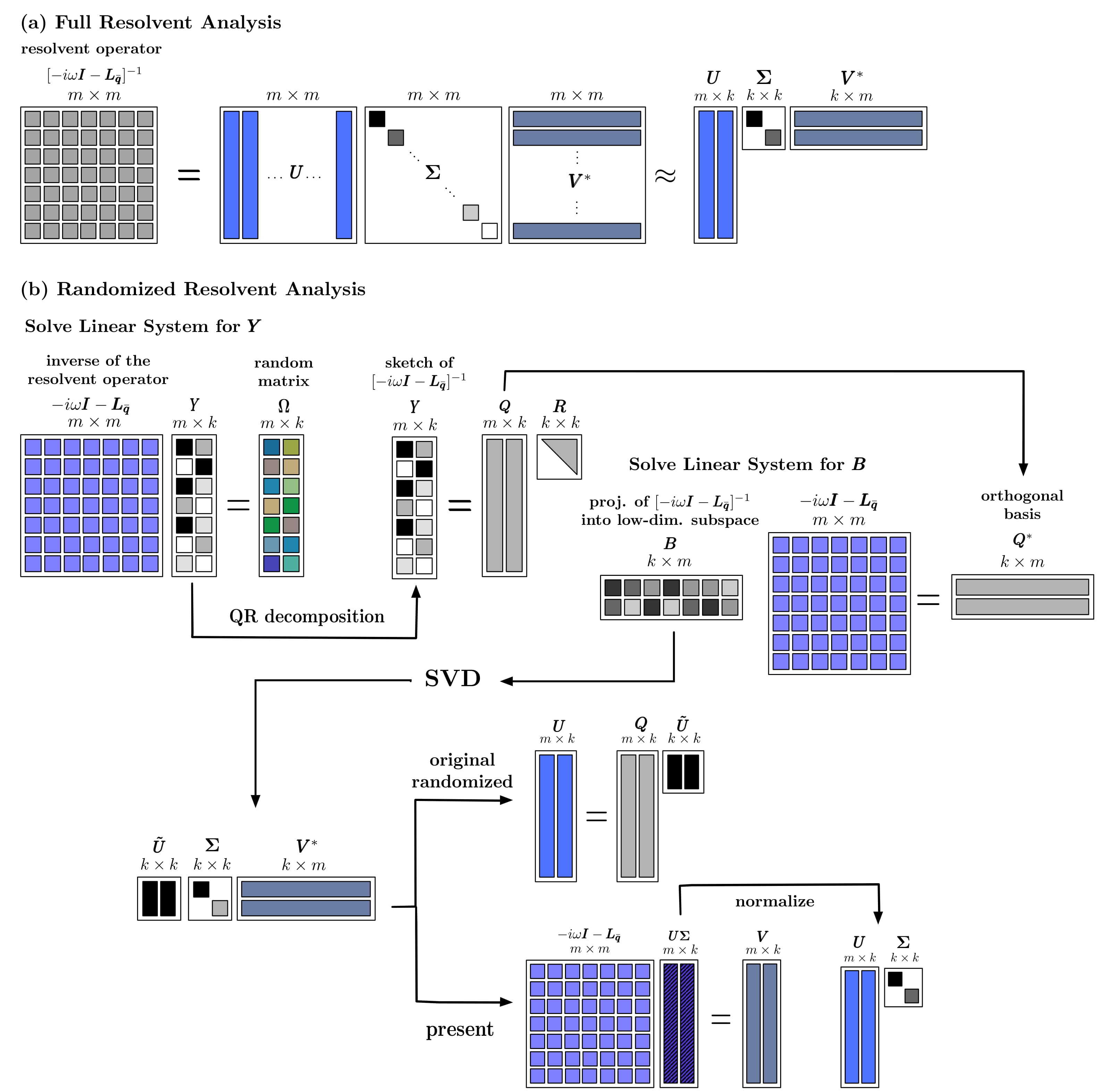}
    \caption{Schematics of the (a) full resolvent and (b) randomized resolvent analyses. For the full resolvent analysis, a direct SVD is applied. For the randomized resolvent analysis, the orthogonal basis $\bm{Q}$ is computed to project the operator into the low-dimensional subspace, where the SVD is performed on the reduced matrix $\bm{B}$. Two options to recover the left singular vectors and singular values are provided, using the original randomized approach \cite{Halko:SIAMReview11} and the present implementation. Orthogonal vectors are represented by long bars.}
    \label{fig:blockdiagram}
\end{figure}

Both full resolvent and randomized resolvent analyses are shown schematically in figure \ref{fig:blockdiagram}. 
Notice that we are not interested in all singular values and vectors of $\left[ -i\omega\bm{I} - \bm{L}_{\bar{\pmb{q}}} \right]^{-1}$, but only in a few subset of the largest $\sigma_j$ and their corresponding $\hat{\bs{u}}_j$ and $\hat{\bs{v}}_j$. In the randomized resolvent analysis, we can approximate a low-rank representation of it using $\left[ -i\omega\bm{I} - \bm{L}_{\bar{\pmb{q}}} \right]$. Figure \ref{fig:blockdiagram} shows an adaptation of the procedure from \citet{Halko:SIAMReview11} in order to compute the largest singular values of the resolvent without performing its inverse. In the randomized resolvent analysis, we solve a linear system with $\left[ -i\omega\bm{I} - \bm{L}_{\bar{\pmb{q}}} \right]$, the columns of the random matrix $\bs{\Omega}$ form the right-hand side and the sketch columns of $\bm{Y}$ are the unknowns. By doing so, we sketch $\left[ -i\omega - \bm{L}_{\bar{\pmb{q}}} \right]^{-1}$ without finding the actual inverse matrix. The same procedure is performed to project the matrix to the low-dimensional subspace. The matrices are re-arranged in a way that the projection is performed using $\left[ -i\omega\bm{I} - \bm{L}_{\bar{\pmb{q}}} \right]$, but results in the low-dimensional projection of $\left[ -i\omega - \bm{L}_{\bar{\pmb{q}}} \right]^{-1}$ instead. Figure \ref{fig:blockdiagram} also illustrates the procedures for recovering the left singular vectors and the singular values from the original algorithm \citep{Halko:SIAMReview11} and from the present implementation. 

{\small
\begin{algorithm}
\caption{Randomized Resolvent Analysis}\label{algorithm}
\begin{algorithmic}[1]
	\SetAlgoNoLine
	\DontPrintSemicolon
	\SetKwFunction{FMain}{randomized$\_$resolvent}
	\SetKwProg{Fun}{Function}{:}{}
	\REQUIRE Discrete linear operator $\bm{L}_{\bar{\pmb{q}}} \in \mathbb{C}^{m\times m}$ \;
	\Fun{\FMain{$\omega$,$k$}}{			
		\nl $\boldsymbol{\Omega} \gets$ \texttt{randn}($m$,$k$)   		\tcp*[r]{Random normal matrix generator. For scaling, see Section \ref{sec:physicsomega}}	   
		$\bm{Y} \gets \left[ -i\omega\bm{I} - \bm{L}_{\bar{\pmb{q}}} \right]  \symbol{92}  \boldsymbol{\Omega}$    \tcp*[r]{Solve linear system for $\bm{Y}$, $\mathcal{O}(m^2k)$}   
		$\left(\bm{Q},\sim \right) \gets \texttt{qr(}\bm{Y}\texttt{,0)}$       									\tcp*[r]{Economy-sized QR decomposition, $\mathcal{O}(mk^2)$}
		$\bm{B} \gets \bm{Q}^* /  \left[ -i\omega\bm{I} - \bm{L}_{\bar{\pmb{q}}} \right] $     						\tcp*[r]{Solve linear system for $\bm{B}$, $\mathcal{O}(m^2k)$}
		$\left( \sim, \sim, \bm{V} \right) \gets \texttt{svd}(\bm{B}\texttt{,`econ')}$  							\tcp*[r]{Reduced SVD decomposition, $\mathcal{O}(mk^2)$}
		$\bm{U}^{\bs{\Sigma}} \gets \left[ -i\omega\bm{I} - \bm{L}_{\bar{\pmb{q}}} \right] \symbol{92} \bm{V}$	\tcp*[r]{solve linear system to recover $\bm{U} {\bs{\Sigma}}$, $\mathcal{O}(m^2k)$}
		\For{$j \gets 1$ to $k$}{
   			$\bs{\Sigma}_{j,j} \gets \texttt{norm}(\bm{U}^{\bs{\Sigma}}_{1:m,j},2)$ 							\tcp*[r]{Recover singular values $\bs{\Sigma}$}
    		$\bm{U}_{1:m,j} \gets \bm{U}^{\bs{\Sigma}}_{1:m,j} / \bs{\Sigma}_{j,j}$ 						\tcp*[r]{Normalize $\bm{U}$}
    	}
		$(\bm{U}, \bs{\Sigma}, \bm{\tilde{V}}) \gets \texttt{svd}(\bm{U}^{\bs{\Sigma}}\texttt{,`econ')}$  			\tcp*[r]{(Optional) Recover $\bm{U}$ and $\bs{\Sigma}$, $\mathcal{O}(mk^2)$}
		$\bm{V} \gets \bm{V} \bm{\tilde{V}}$  												\tcp*[r]{(Optional) Recover improved $\bm{V}$, $\mathcal{O}(mk)$}
	\Return ($\bm{U}$,$\bs{\Sigma}$,$\bm{V}$)}
\end{algorithmic}
\end{algorithm}
}

\subsubsection{Oversampling and power iteration schemes}
\label{sec:oversampling}

Randomized algorithms can incorporate two additional procedures to improve performance and accuracy. Namely, they are oversampling \citep{Erichson:JSS19} and power or subspace iterations \citep{Rocklin:SIAMJMAA09, Halko:SIAMReview11, Gu:SIAM15}.  Oversampling sketches the input matrix using $(k + p)$ vectors (with $p$ extra vectors) and increases the low-dimensional subspace to accurately recover a smaller quantity of singular values $k$. For the randomized resolvent analysis, oversampling has the same outcome, in practice, of selecting a larger $k$ and the influence of $k$ will be discussed at the end of section $\ref{sec:results}$. When $k$ becomes large, it should be noticed that the memory consumption increases. Even for large sparse matrices, the sketch matrix and the subsequently reduced matrices that are formed are generally dense, which adds a computational burden. 

The second procedure is the power or subspace iterations. These methods are a powerful tool when the singular values of the matrix decay slowly. For example, this type of spectral behavior appeared in the input-output analysis performed by \citet{Jeun:POF16} for jet flows. The method consists of performing additional iterations after the sketch $\bm{Y}$ is evaluated. It should however be realized that such procedure calls for additional linear solvers, which is the most time consuming operation in Algorithm 1. For power iterations, one must compute the adjoint $\bm{A}\bm{A}^*$, where $\bm{A}^*$ is the Hermitian of $\bm{A}$, $q$ times and solve the linear system $k$ times. For subspace iterations, additional $QR$ decompositions are necessary, and the number of additional linear systems to be solved will be $q$ times $k$. In practice, small values of $q$ improves the accuracy of the results substantially. More information on  the general applicability of the subspace and power iterations are discussed by \citet{Halko:SIAMReview11} and \citet{Erichson:JSS19}. In section \ref{sec:physicsomega}, we present another option for improving accuracy of the overall technique in a computationally inexpensive manner by constructing a physics-informed random test matrix $\bs{\Omega}$.


\subsubsection{Test matrix $\bs{\Omega}$}
\label{sec:test}

The standard choice for the test matrix $\bs{\Omega}$ is the random matrix generated with a normal distributions. Such matrix is known to present excellent performance and accuracy \cite{Tropp:SIAMMAA17}.  For some cases, especially for very large matrices or when the singular values present slow decay, larger values of $k$ may be necessary to better approximate the matrix in the low-dimensional subspace. When large values of $k$ are used, orthonormalization of the columns of the test matrix can be considered to improve numerical stability \citep{Demmel:NM07,Halko:SIAMReview11}.  The test matrix can also be generated using a Rademacher distribution \citep{Clarkson:ACM07}. It is also possible to build an ultrasparse matrix with Rademacher distribution in the non-sparse entries which allows for the control of cost, stability and reliability in the operations \citep{Tropp:SIAMMAA17,Clarkson:ACM07}.

When randomized SVD is applied, there is no a priori knowledge of the structure of the matrix. However, in the present application, we know how the resolvent operator is constructed. This theoretical insight can be used to build a random test matrix that outperforms the standard normal distribution matrix. We later propose a physics-informed test matrix $\bs{\Omega}$ that can focus our sketching operation for regions of physical importance. In our application, the dominant directions are related to regions with the presence of high shear. The results from this approach will be discussed in Section \ref{sec:physicsomega}.

\section{Randomized resolvent analysis of turbulent post-stall flow}
\label{sec:example}

We demonstrate the use of randomized resolvent analysis on turbulent flow over a NACA 0012 airfoil.  In this example, the randomized resolvent analysis will be applied a resolvent operator of size $m \times m$, where $m \simeq 7\times10^5$, to reveal the dominant gain and modal structures with a thin sketching matrix having as little as $k = 10$ columns.  The convergence of the gain and resolvent modes will also be reported with respect to the size of the sketching matrix.  Influence of the ratio between the first and the second singular values of the resolvent operator will also be examined.

\subsection{Problem setup}
\label{sec:setup}

We consider the spanwise-periodic turbulent flow over a NACA 0012 airfoil at an angle of attack of $9^\circ$, a chord-based Reynolds number of $Re_{L_c} \equiv v_\infty L_c/\nu_\infty = 23,000$ and a free stream Mach number of $M_\infty \equiv v_\infty/a_\infty = 0.3$.  Here, $v_\infty$ is the free-stream velocity, $L_c$ is the chord length, $a_\infty$ is the free-stream sonic speed, and $\nu_\infty$ is the kinematic viscosity.  The time- and spanwise-averaged turbulent flow is considered as the base flow for the full and randomized resolvent analyses.  For this 2D base flow, we adopt the bi-global setting that decomposes $\pmb{q}'$ into spanwise Fourier modes with the wavenumber $k_z$.

\begin{figure}[t]
    \vspace{0.05in}
    \centering
    \begin{overpic}[width=0.95\textwidth]{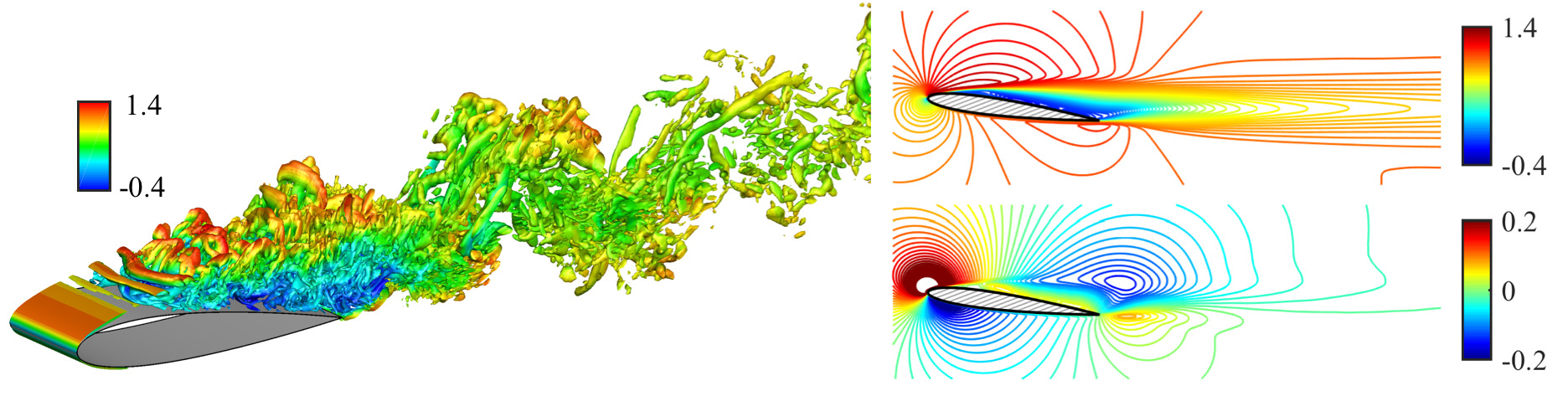}
		\put(85, 23){\small{$\bar{v}_x/v_\infty$}}
		\put(85, 10.5){\small{$\bar{v}_y/v_\infty$}}
		\put( 3, 21){\small{$v_x/v_\infty$}}
	\end{overpic}
    \caption{The instantaneous (left) and time/spanwise-averaged (right) flows over a NACA 0012 airfoil at $Re_{L_c} = 23,000$. The instantaneous flow visualization shows the isosurface of $Q$-Criterion ($QL_c^2/v_\infty^2 = 50$) colored by the instantaneous streamwise velocity.}
    \label{fig:baseflow}
\end{figure}

To obtain the base flow, large-eddy simulation (LES) is performed using a finite-volume compressible flow solver {\it{CharLES}} \citep{Khalighi:AIAA11,Bres:AIAAJ2017}, which is second-order accurate in space and third-order accurate in time.  Vremen's sub-grid scale model \citep{Vreman:PF04} is utilized in the LES.  The LES is performed on a C-shaped mesh with the domain extent of $x/L_c \in [-19, 26]$, $y/L_c \in [-20, 20]$ and $z/L_c \in [-0.1, 0.1]$ in the streamwise, transverse and spanwise direction, respectively, with the airfoil leading edge at $x/L_c = y/L_c = 0$.  Dirichlet boundary condition is specified at the far-field boundary as $(\rho, v_x, v_y, v_z, T) = (\rho_\infty, v_\infty, 0, 0, T_\infty)$, where $\rho$ is the density, $v_x$, $v_y$ and $v_z$ are respectively the streamwise, transverse and spanwise velocity, and $T$ is the temperature.  Over the airfoil, the no-slip adiabatic boundary condition is prescribed.  Along the outlet boundary, a sponge layer \citep{Freund:AIAAJ97} is applied with a running-averaged state being the target state.
The simulation has been validated with respect to the time-averaged pressure, lift and drag over the airfoil.  
The turbulent separated flow over the airfoil is visualized in figure \ref{fig:baseflow}.  The visualization of the instantaneous flow shows the laminar separation from the leading edge.  
We have found that the shear layer physics dominates the pseudospectral behavior of the linearized Navier--Stokes operator, as shear is the main source of nonnormality in the operator.
Further details regarding the computational setup, flow physics, and resolvent analysis based flow control of this setup are reported in \citet{Yeh:JFM19}.

The full and randomized resolvent analyses are performed on a separate mesh from that used in the LES.  This mesh has a 2D rectangular domain with the extent of $x/L_c \in [-15, 16]$ and $y/L_c \in [-12, 12]$, comprising approximately $0.15$ million cells.  Compared to the LES mesh, the mesh for resolvent analysis is coarser over the airfoil and in the wake, but is much finer in the upstream of the airfoil in order to resolve the forcing mode structures.  The time- and spanwise-averaged flow $\bar{\boldsymbol{q}}$
obtained from  LES is interpolated onto this mesh.  
At the far-field boundary and over the airfoil, Dirichlet conditions are set for density and velocities and Neumann condition is prescribed for pressure in $\pmb{q}'$. At the outlet boundary, Neumann condition is set for all flow variables.  With these boundary conditions for $\pmb{q}'$ and the base flow $\bar{\boldsymbol{q}}$, we construct the linearized Navier--Stokes operator $\bm{L}_{\bar{\boldsymbol{q}}}(k_z)$ for a chosen $k_z$.  
The size of $\bm{L}_{\bar{\boldsymbol{q}}}$ and the resolvent operator is approximately $0.75$ million $\times$ $0.75$ million.

For this large operator, we summarize in Table \ref{tab:comp} the computational costs of performing resolvent analysis using the Krylov-based Arnoldi-iteration method with a range of parameter setups (i.e., number of singular values ($n_{\text{ev}}$), Krylov subspace dimension ($\text{dim}(\mathcal{S})$), and tolerance) and compare them with those for the present randomized algorithm.  The former was conducted by simply calling the {\sf svds} command in MATLAB.  It requires almost $80$ gigabytes of memory and takes approximately $30$ to $70$ minutes (single-core) for each SVD.  The high-memory demand necessitates the use of high performance computing resource to conduct the full resolvent analysis.  In contrast, the randomized resolvent approach (Algorithm 1) achieves significant reductions in computational time and memory consumption.  The present method only requires a third of the memory usage of the Arnoldi-iteration and cuts down the computational time by an order of magnitude.  We also note that, in Algorithm 1, the linear systems solvers are the operations with higher computational cost.  Since all the three linear systems solvers are conducted for the same operator, the LU decomposition of $\left[ -i\omega\bm{I} - \bm{L}_{\bar{\pmb{q}}} \right]$ is performed in the beginning of the algorithm and is passed through the three solvers.  This decomposition becomes the main source of the memory consumption.

\begin{table}
\begin{center}

\begin{tabular}{p{0.7in}p{0.7in}p{0.8in}p{0.8in}p{0.8in}}
\multicolumn{5}{l}{\bf Iteratively restarted Arnoldi method (MATLAB {\sf svds})}\\
\hline
$n_{\text{ev}}$ & $\text{dim}(\mathcal{S})$ & tolerance & time (sec) & memory  \\ 
\hline
$10$ & $30$ & 1E-14 & 4185 & 78.6 GB  \\
$5$  & $15$ & 1E-14 & 2764 & 78.6 GB  \\
$2$  & $6$  & 1E-14 & 1486 & 78.6 GB  \\
$10$ & $30$ & 1E-05 & 2245 & 78.6 GB  \\
$10$ & $15$ & 1E-14 & 4194 & 78.6 GB  \\ \hline
~\\

\multicolumn{5}{l}{\bf Randomized resolvent (present)}\\
\hline
$n_{\text{ev}}$ & $k$ & & time (sec) & memory \\ 
\hline
$2$  & $2$	&	& 354 & 28.2 GB  \\
$5$  & $5$	&	& 462 & 28.4 GB  \\
$10$ & $10$	&   & 615 & 28.8 GB  \\ \hline

\end{tabular}

\caption{\label{tab:comp}Comparison of the computational time and memory consumption for the implicitly restarted Arnoldi iteration ({\sf svds} in MATLAB) with the present randomized approach for different parameter setups with the number of singular values ($n_{\text{ev}}$), Krylov subspace dimension ($\text{dim}(\mathcal{S})$), and tolerance. }
\end{center}
\end{table}

\subsection{Results}
\label{sec:results}

We perform the full and randomized resolvent analyses for spanwise wavenumbers of $k_zL_c = 0$ and $20\pi$.  Since the base flow is found to be unstable \citep{Yeh:JFM19}, the finite-time approach is adopted with $ v_\infty / \beta L_c = 3$ to ensure that the resolvent analysis is performed on a shorter time scale than that associated with the leading growth rate of the instability.  Initially, for the randomized analysis, we consider $k = 10$ for the width of the test matrix $\bs{\Omega}$ with random Gaussian distribution. Later in Section \ref{sec:physicsomega}, we show that this value of $k$ can be further reduced without compromising accuracy.  

\begin{figure}
\footnotesize
 	\begin{tabular}{	>{\centering\arraybackslash} m{0.05in}  
						>{\centering\arraybackslash} m{0.25in}  
						>{\centering\arraybackslash} m{1.51in}
						>{\centering\arraybackslash} m{1.51in}
						c
						>{\centering\arraybackslash} m{1.51in}
						>{\centering\arraybackslash} m{1.51in}}
	&&\multicolumn{2}{c}{\bf Full resolvent analysis}\vspace{-0.00in}&&\multicolumn{2}{c}{\bf Randomized resolvent analysis}\\
	\cline{3-4}\cline{6-7}\\
			&\vspace{-0.06in}$St$
			&\vspace{-0.06in}Response mode $\hat{\pmb{u}}_1^\text{full}$
			&\vspace{-0.06in}Forcing mode $\hat{\pmb{v}}_1^\text{full}$ 
			& 
			&\vspace{-0.06in}Response mode $\hat{\pmb{u}}_1^\text{rand}$
			&\vspace{-0.06in}Forcing mode $\hat{\pmb{v}}_1^\text{rand}$\\
    \hline
    		&   $1$
				\vspace{+0.04in}
			&	\vspace{+0.04in} \includegraphics[width=1.50in]{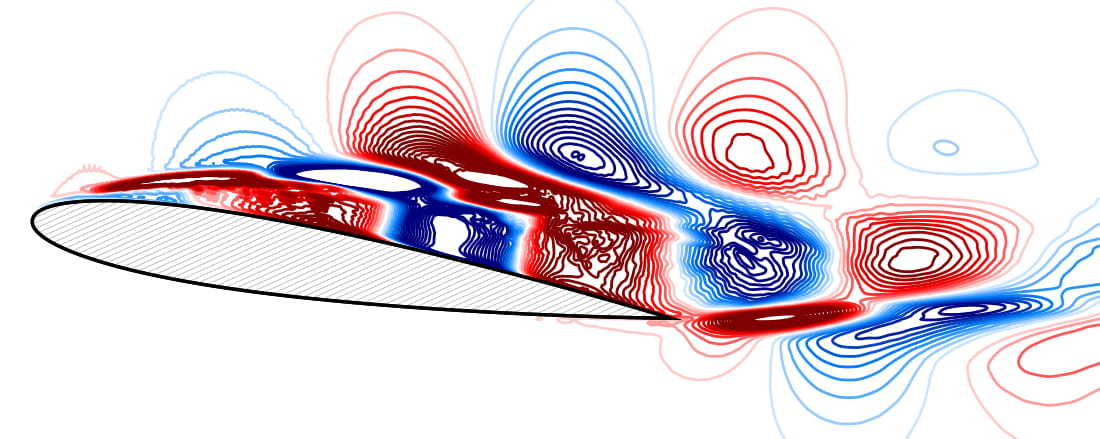}
			&	\vspace{+0.04in} \includegraphics[width=1.50in]{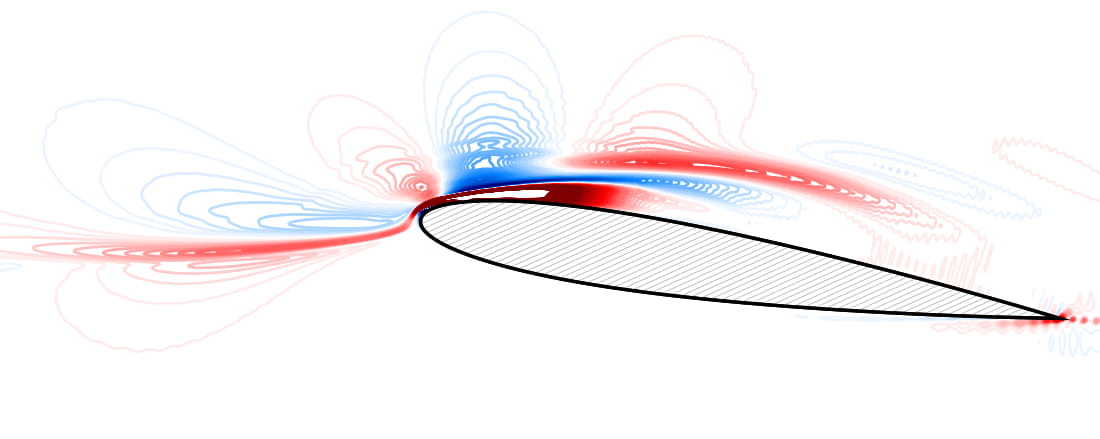}
			&
			&	\vspace{+0.04in} \includegraphics[width=1.50in]{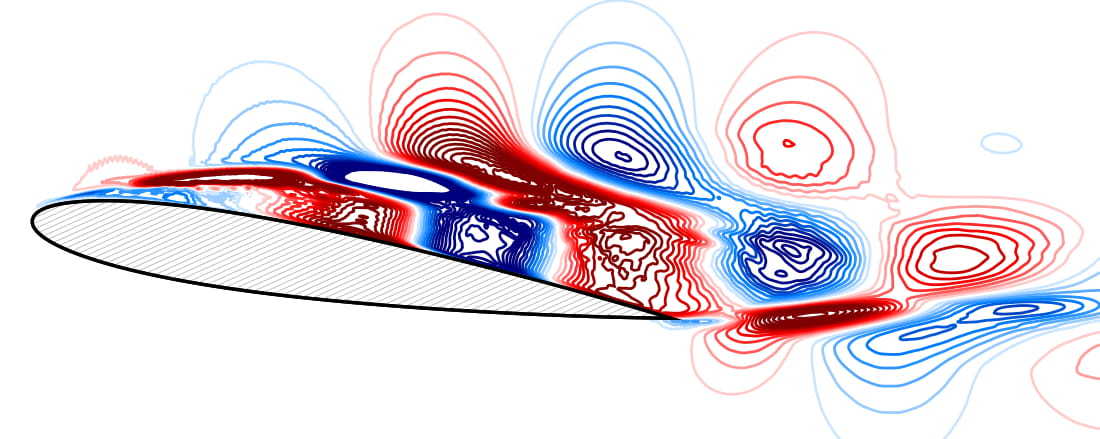}
			&	\vspace{+0.04in} \includegraphics[width=1.50in]{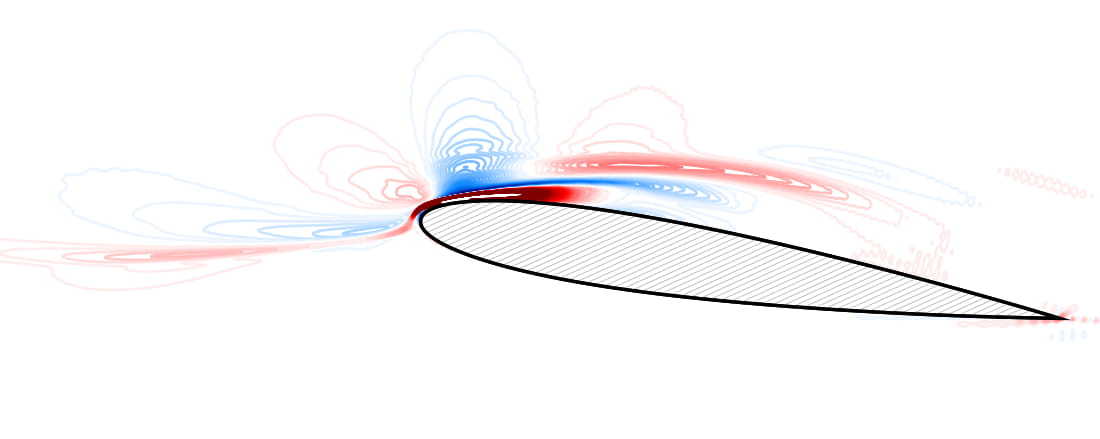}\\
	\multirow{2}{*}{\rotatebox{90}{$k_zL_c = 0$}}\vspace{+0.04in}
    		&   $3$
				\vspace{+0.04in}
			&	\vspace{+0.04in} \includegraphics[width=1.50in]{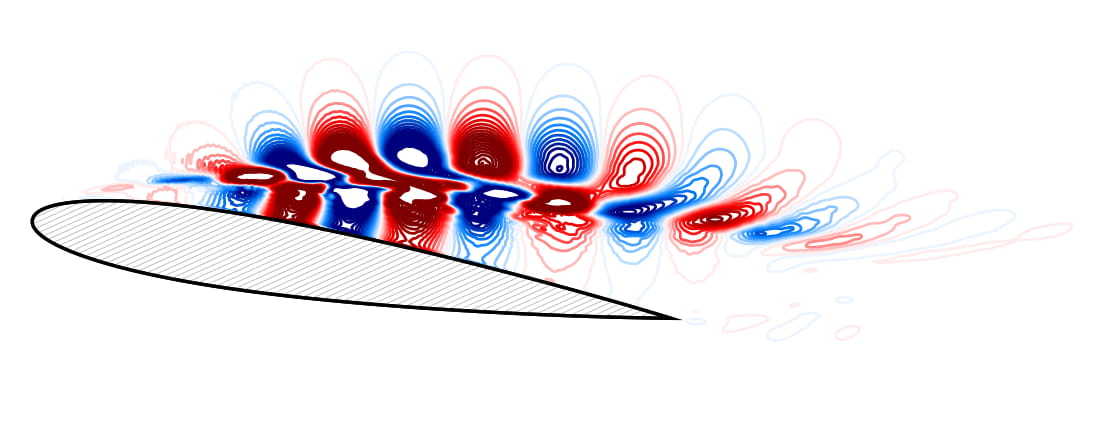}
			&	\vspace{+0.04in} \includegraphics[width=1.50in]{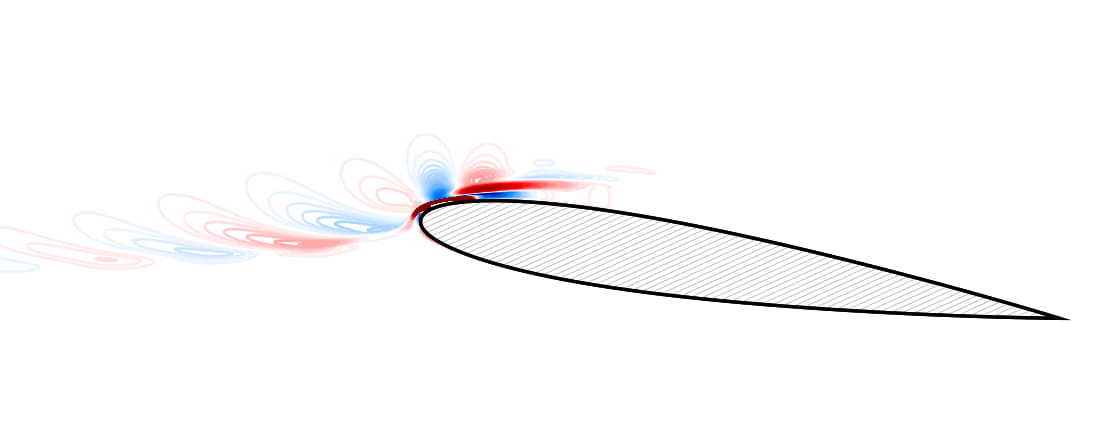}
			&
			&	\vspace{+0.04in} \includegraphics[width=1.50in]{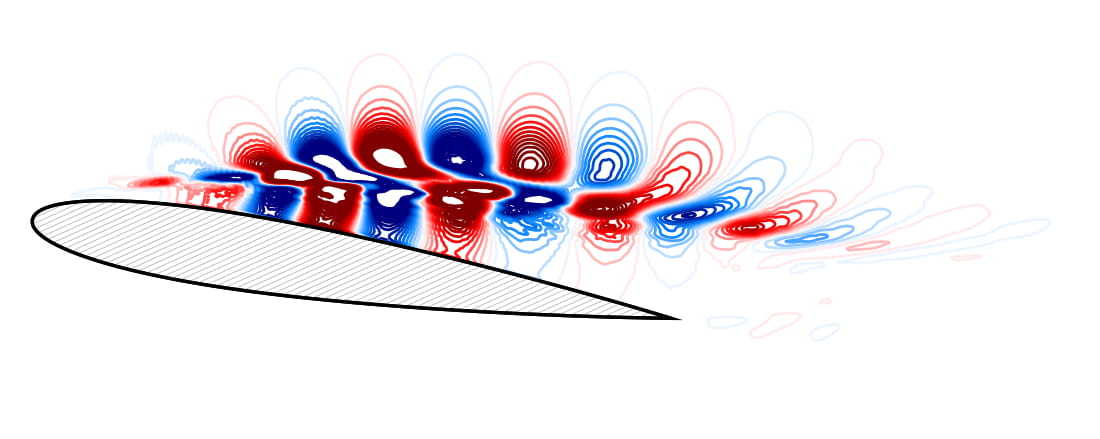}
			&	\vspace{+0.04in} \includegraphics[width=1.50in]{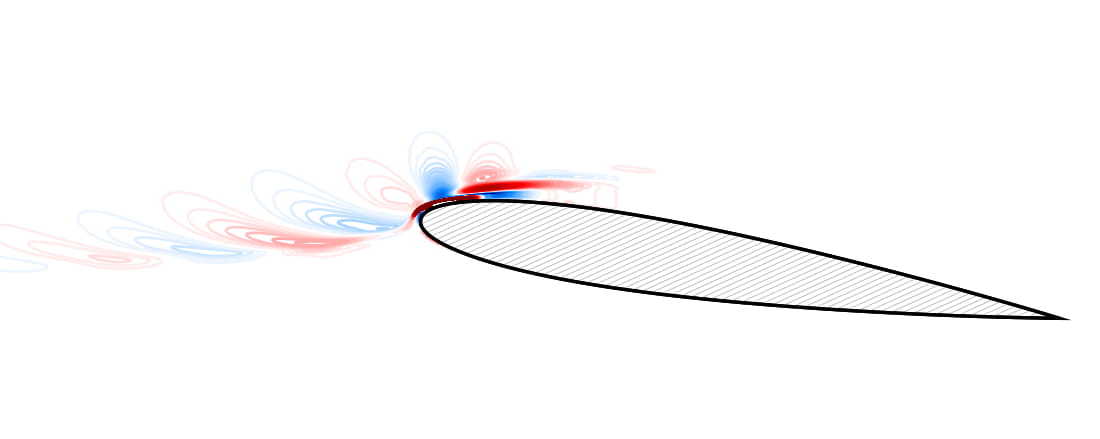}\\
    		&   $6$
				\vspace{+0.04in}
			&	\vspace{+0.04in} \includegraphics[width=1.50in]{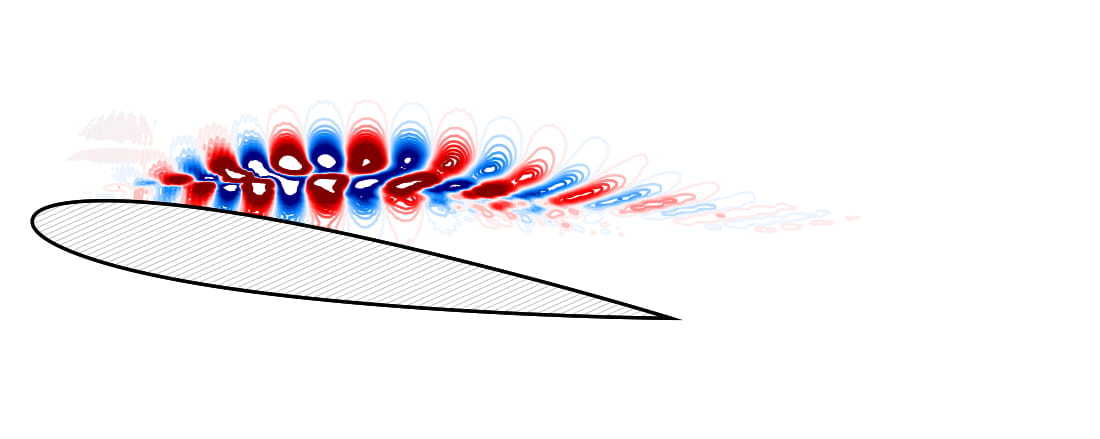}
			&	\vspace{+0.04in} \includegraphics[width=1.50in]{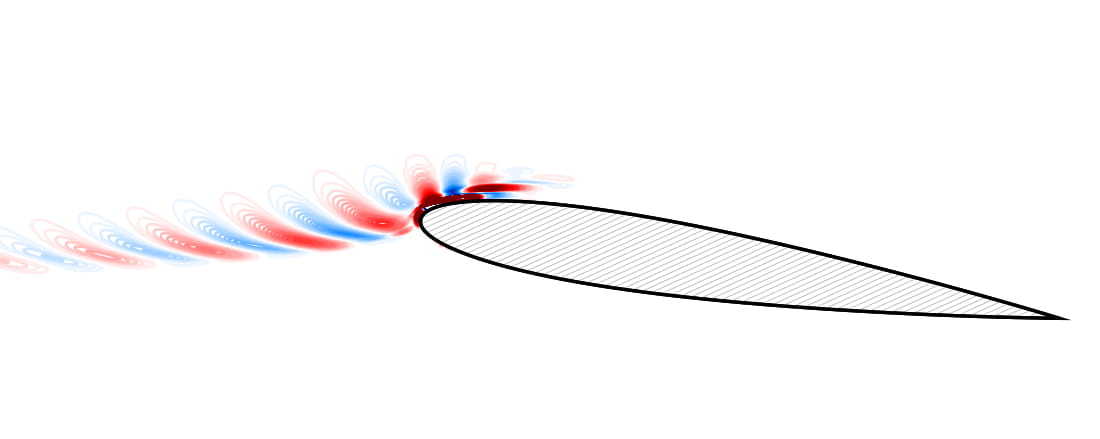}
			&
			&	\vspace{+0.04in} \includegraphics[width=1.50in]{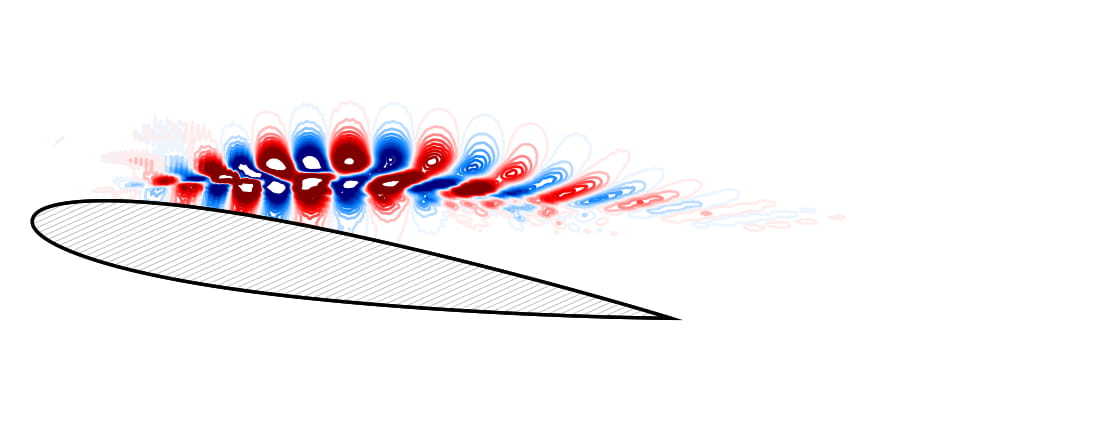}
			&	\vspace{+0.04in} \includegraphics[width=1.50in]{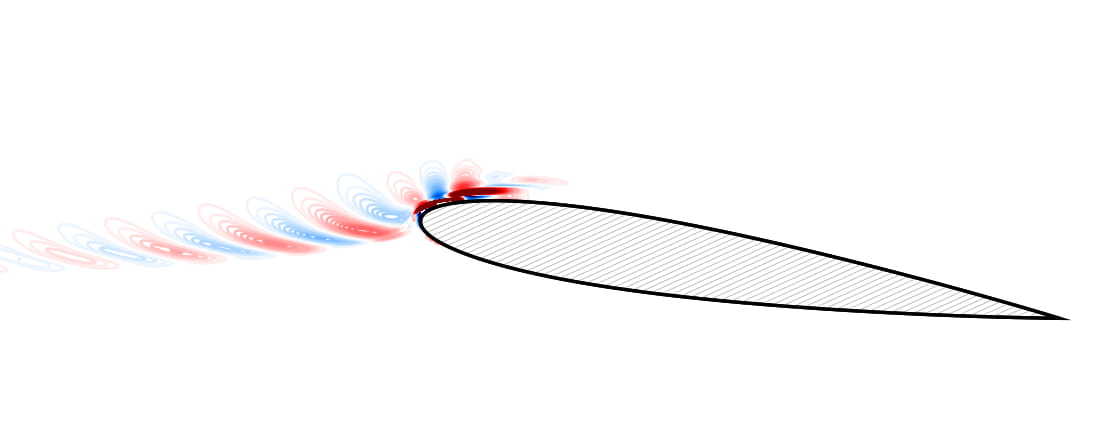}\\
    		&   $12$
			&	\vspace{+0.04in} \includegraphics[width=1.50in]{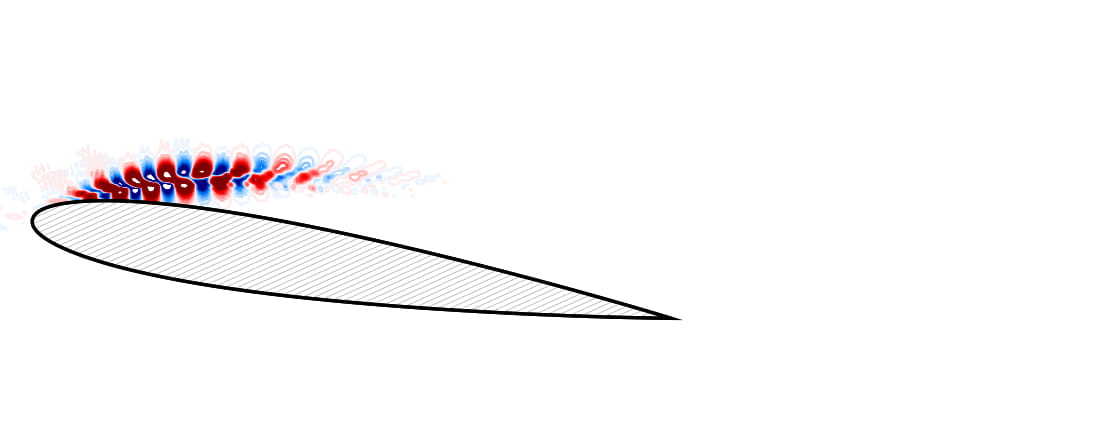}
			&	\vspace{+0.04in} \includegraphics[width=1.50in]{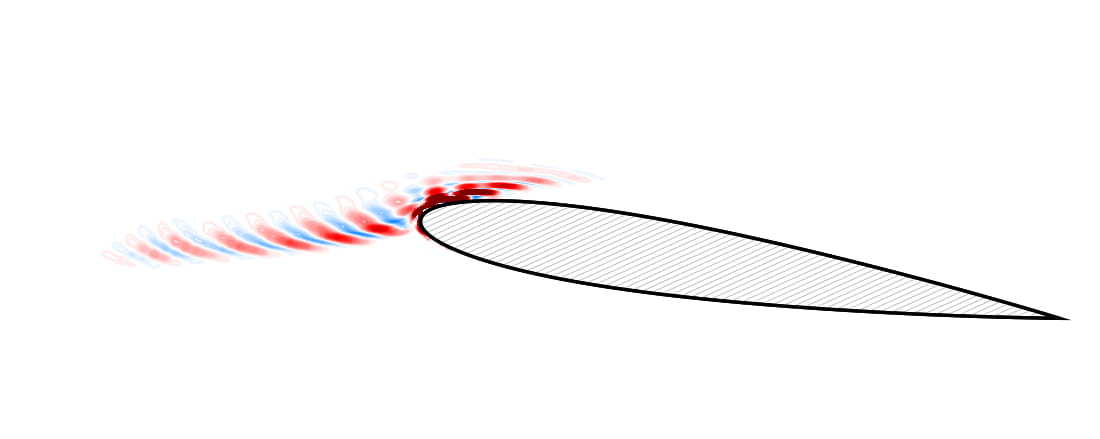}
			&
			&	\vspace{+0.04in} \includegraphics[width=1.50in]{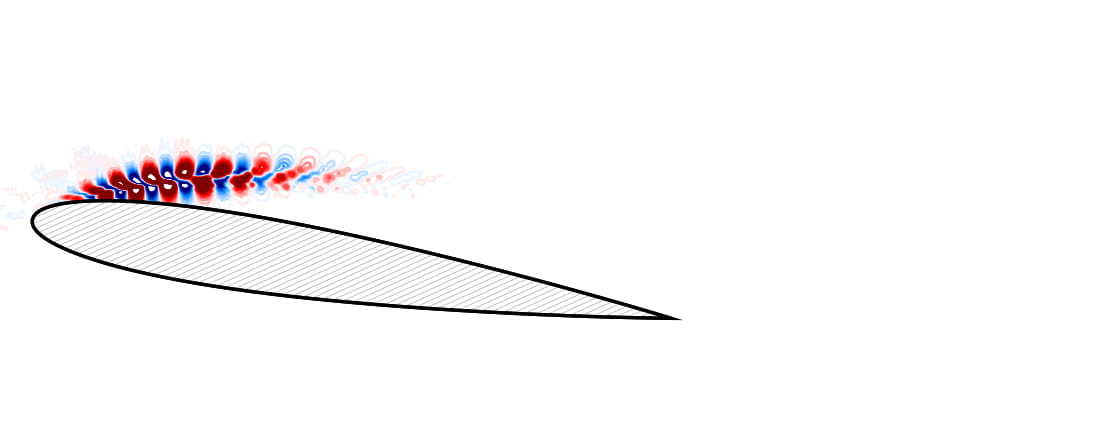}
			&	\vspace{+0.04in} \includegraphics[width=1.50in]{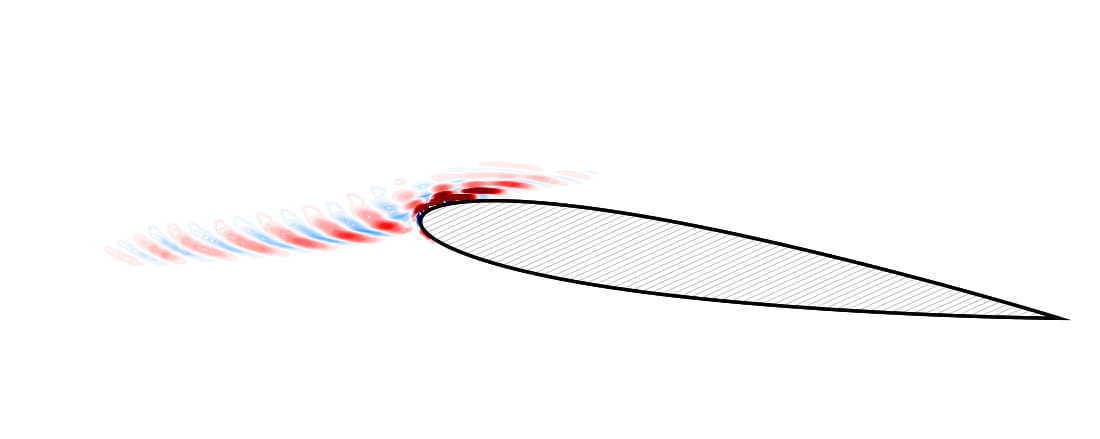}\\
	\hline
    		&   $1$
				\vspace{+0.04in}
			&	\vspace{+0.04in} \includegraphics[width=1.50in]{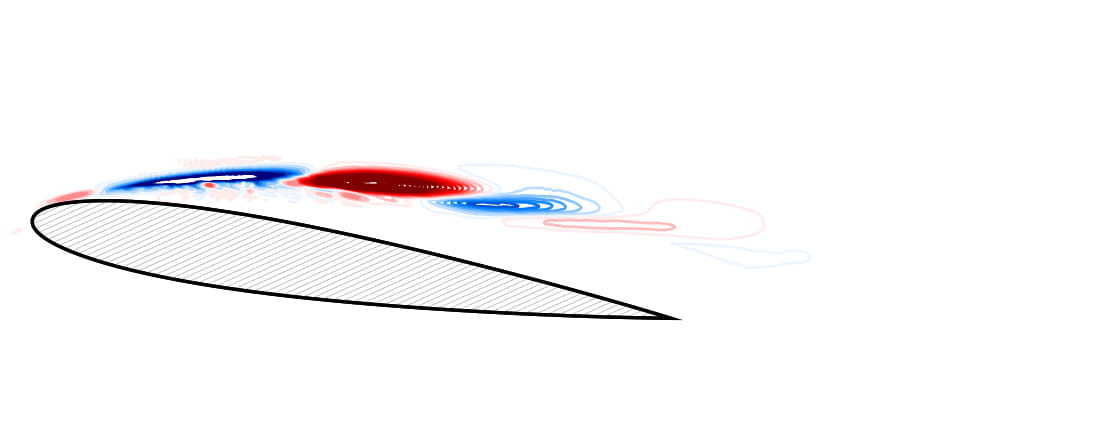}
			&	\vspace{+0.04in} \includegraphics[width=1.50in]{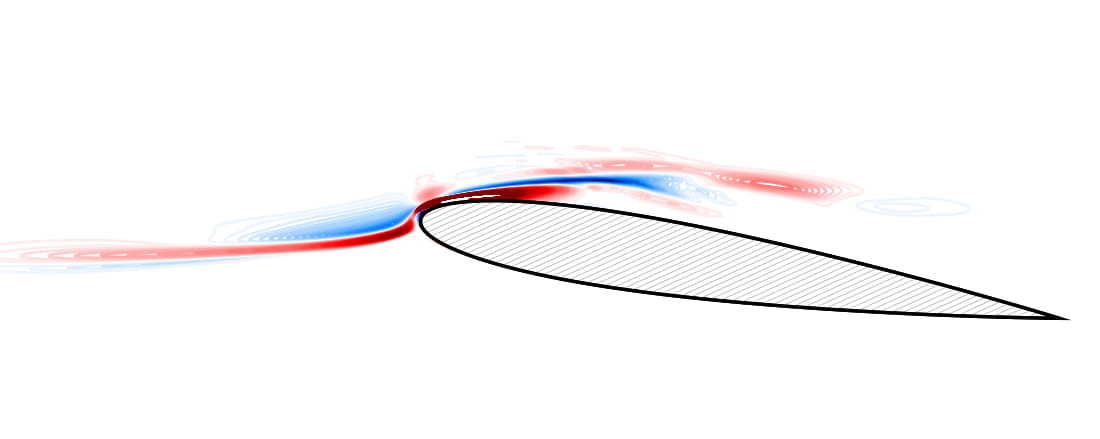}
			&
			&	\vspace{+0.04in} \includegraphics[width=1.50in]{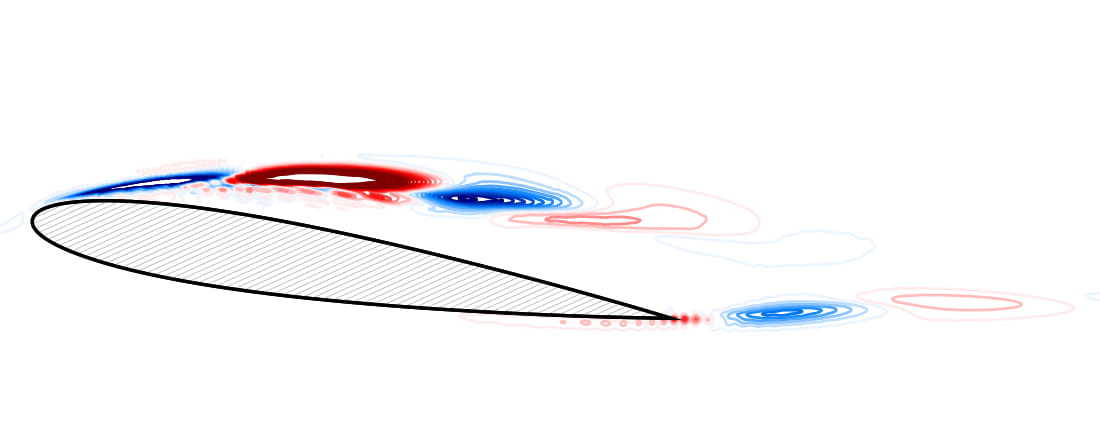}
			&	\vspace{+0.04in} \includegraphics[width=1.50in]{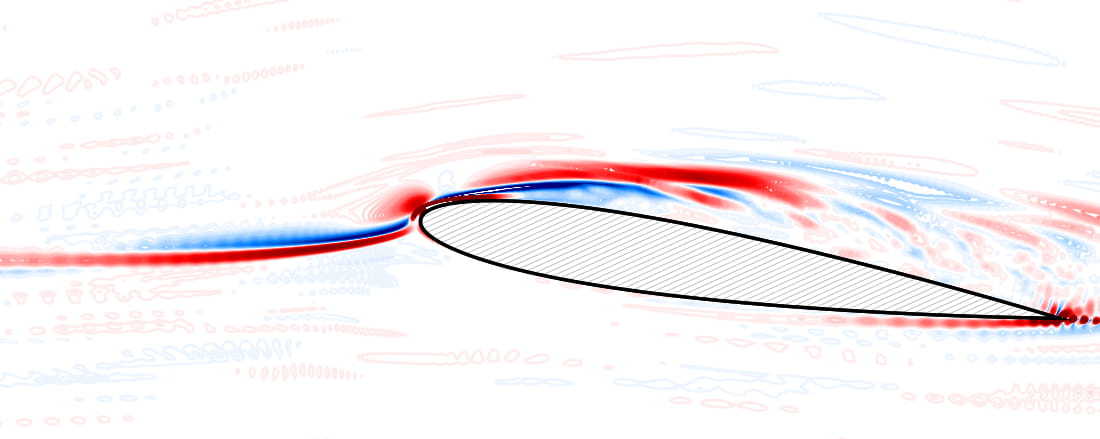}\\
	\multirow{2}{*}{\rotatebox{90}{$k_zL_c = 20\pi$}}\vspace{+0.04in}
    		& $3$
				\vspace{+0.04in}
			&	\vspace{+0.04in} \includegraphics[width=1.50in]{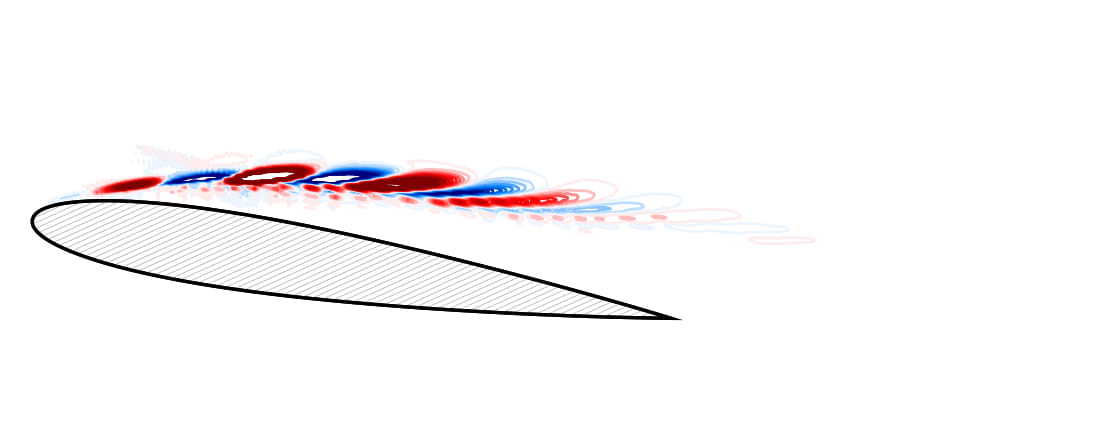}
			&	\vspace{+0.04in} \includegraphics[width=1.50in]{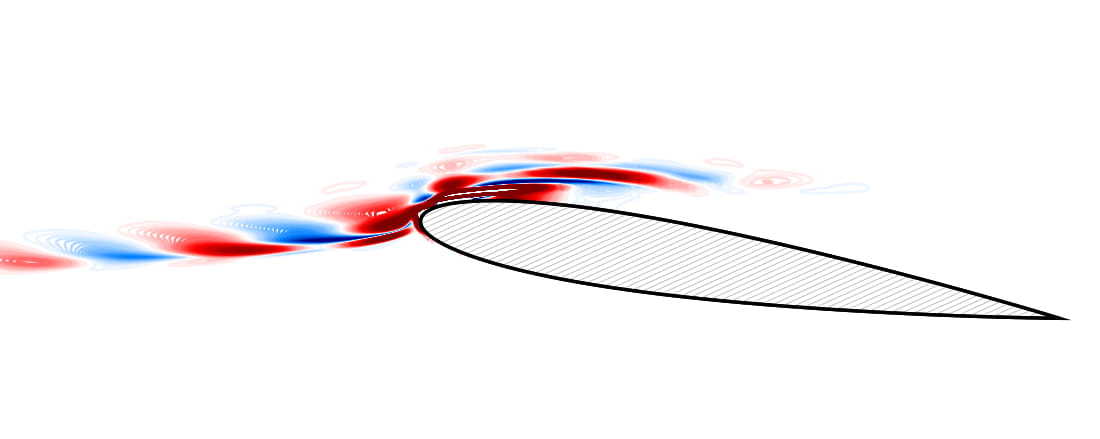}
			&
			&	\vspace{+0.04in} \includegraphics[width=1.50in]{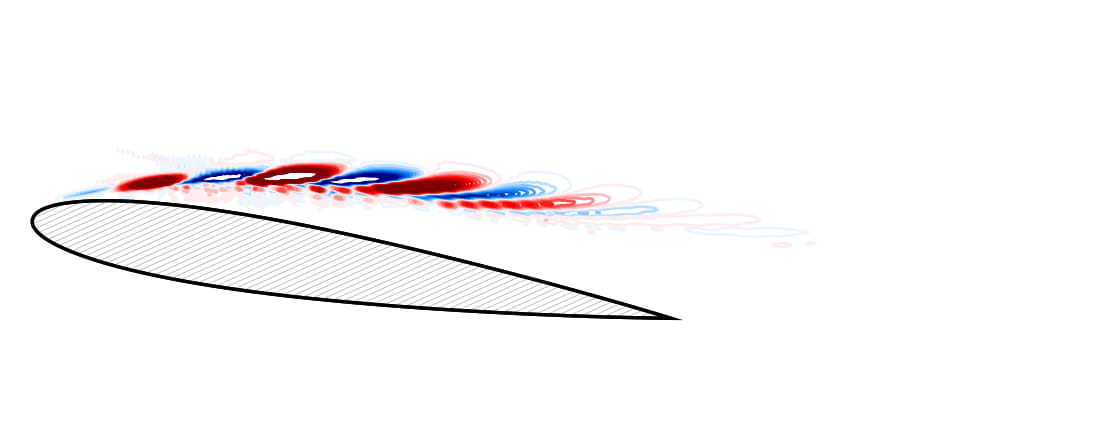}
			&	\vspace{+0.04in} \includegraphics[width=1.50in]{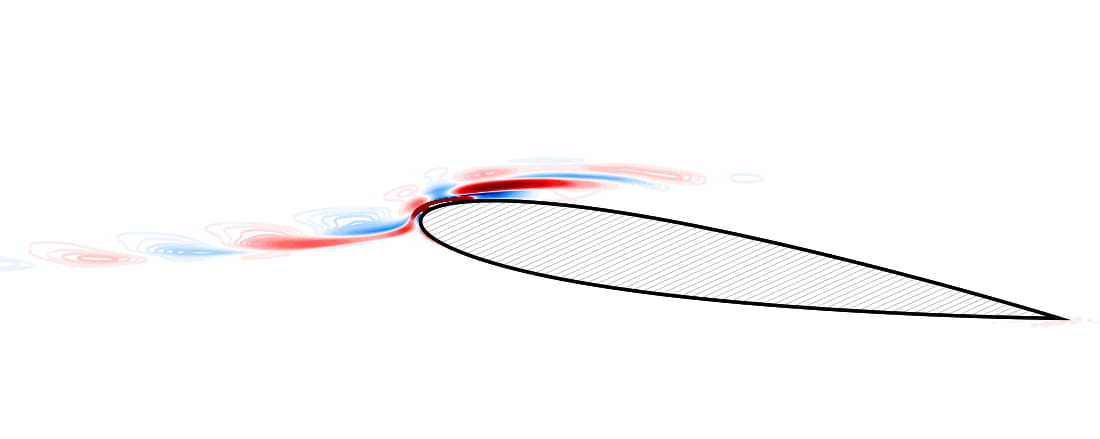}\\
    		& $6$
				\vspace{+0.04in}
			&	\vspace{+0.04in} \includegraphics[width=1.50in]{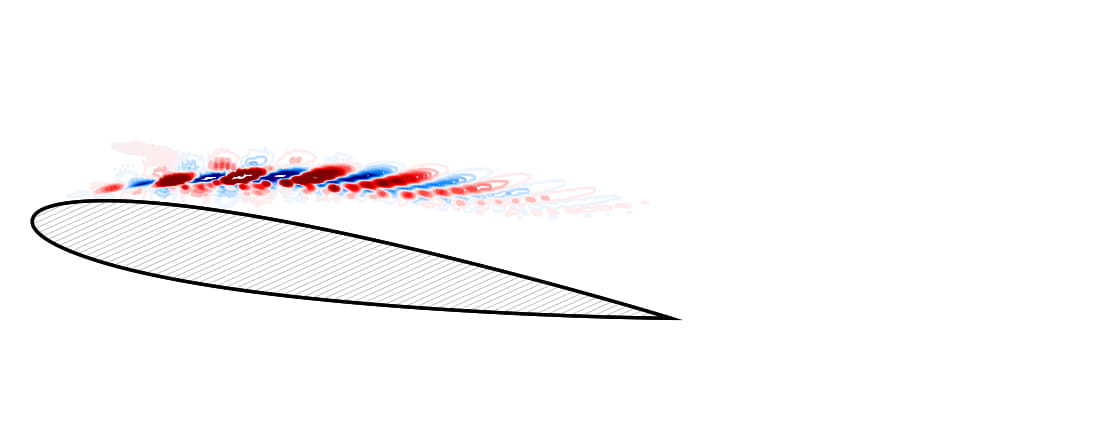}
			&	\vspace{+0.04in} \includegraphics[width=1.50in]{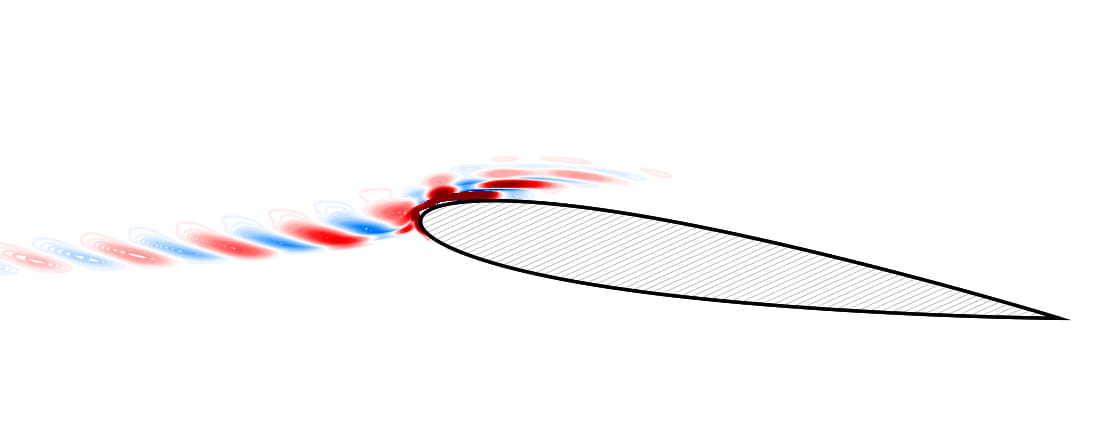}
			&
			&	\vspace{+0.04in} \includegraphics[width=1.50in]{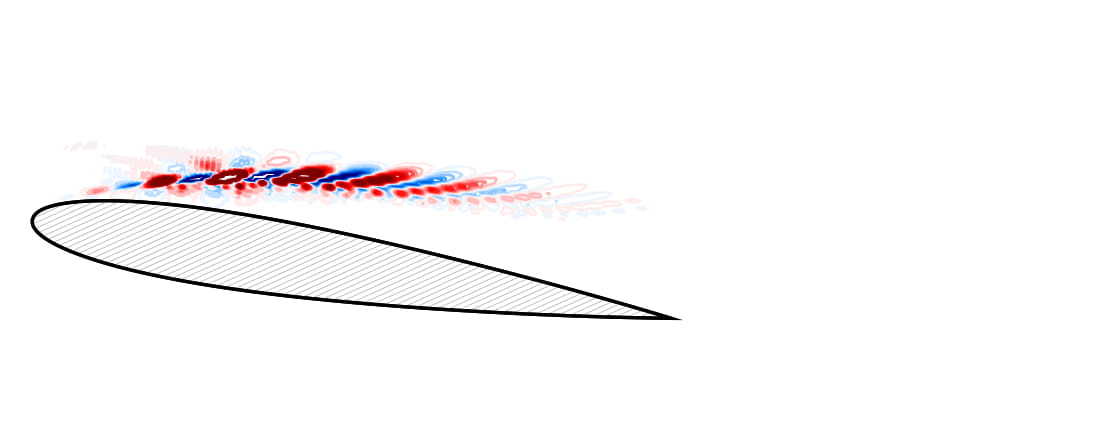}
			&	\vspace{+0.04in} \includegraphics[width=1.50in]{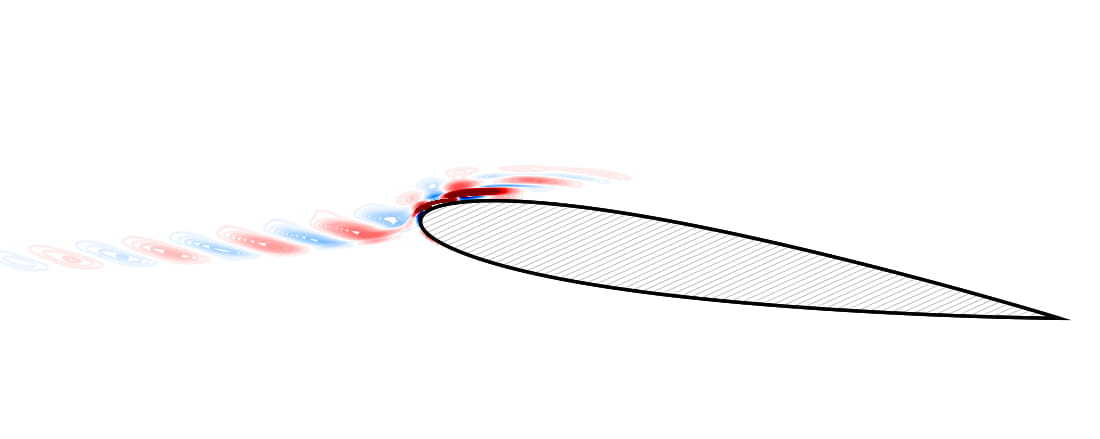}\\
    		& $12$
				\vspace{+0.04in}
			&	\vspace{+0.04in} \includegraphics[width=1.50in]{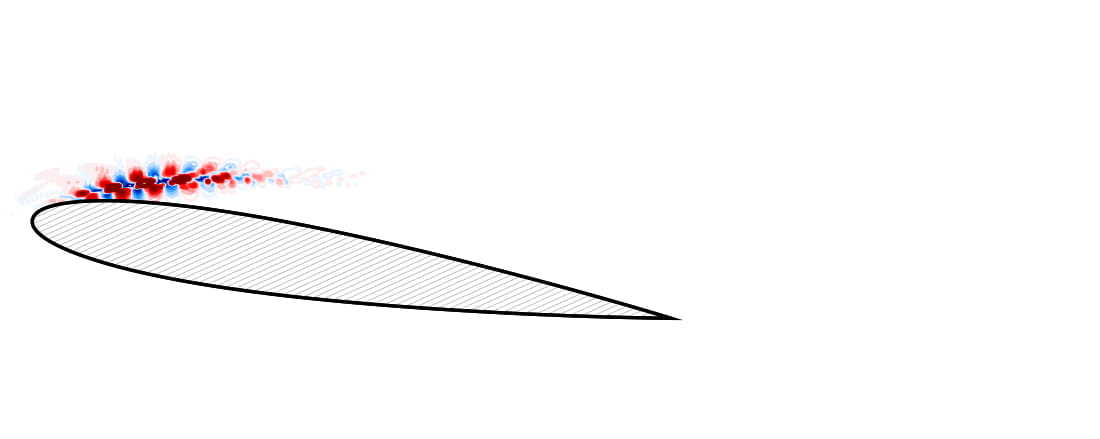}
			&	\vspace{+0.04in} \includegraphics[width=1.50in]{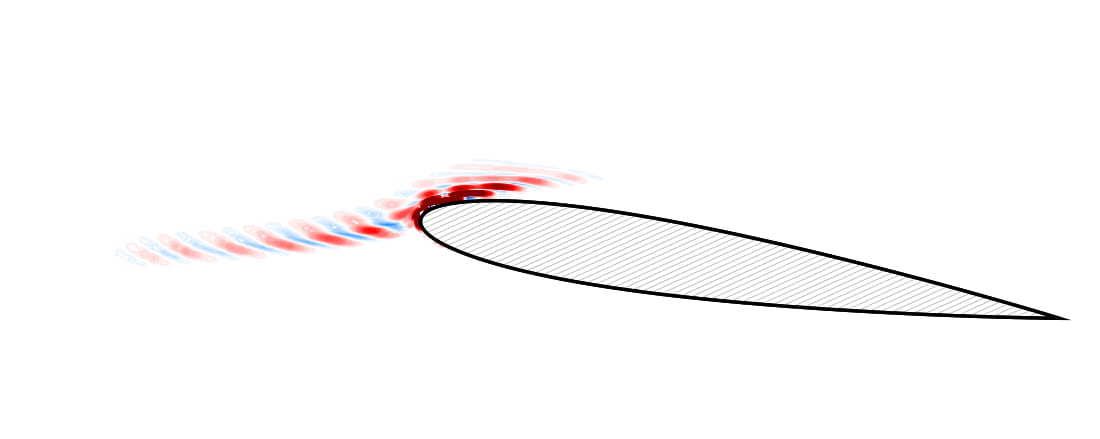}
			&
			&	\vspace{+0.04in} \includegraphics[width=1.50in]{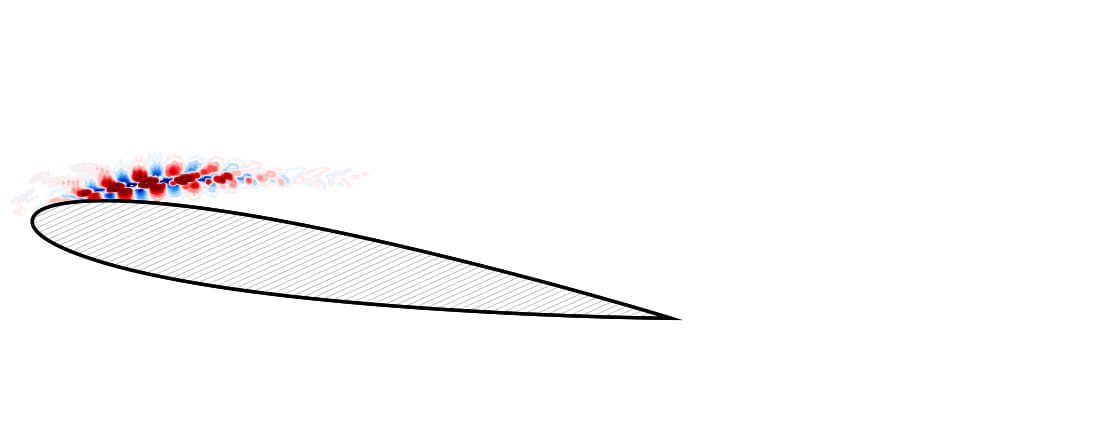}
			&	\vspace{+0.04in} \includegraphics[width=1.50in]{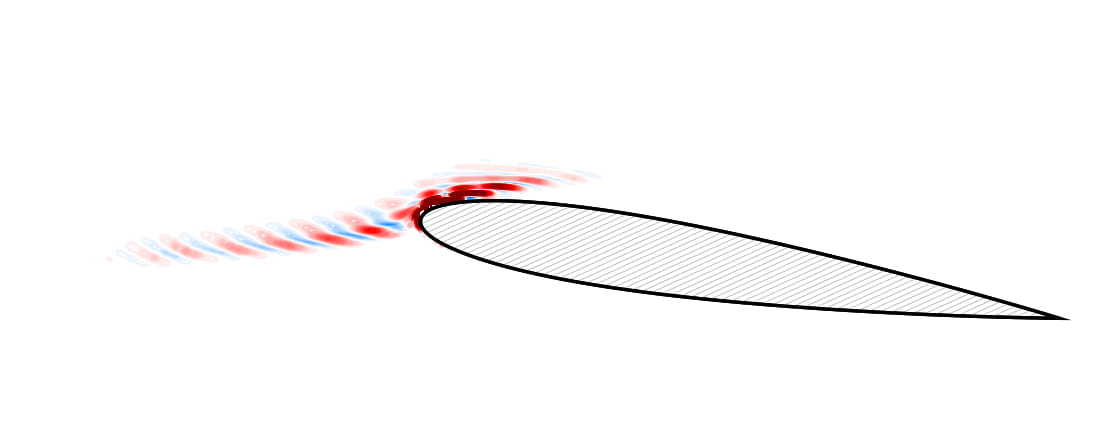}\\
	\end{tabular}
\caption{The leading response ($\hat{\pmb{u}}_1$) and forcing ($\hat{\pmb{v}}_1$) modes from the full and randomized resolvent analyses, using Algorithm 1, for $k_zL_c = 0$ and $20\pi$ at representative frequencies $St$. Modes are visualized with the streamwise velocity component with contour levels of $[-0.6, 0.6]$.} 
\label{fig:plot_mode_avs}
\end{figure}

The leading response and forcing modes obtained from both full and randomized analyses are compared in figure \ref{fig:plot_mode_avs} for representative frequencies $St$ and spanwise wavelength $k_z L_c$.  Although $k/m = 1.3 \times 10^{-5}$, we observe excellent agreement between the modes from the full resolvent analysis and the randomized algorithm. We observe that randomized forcing and response modes are very similar to full resolvent ones. Only at $St = 1$ and $k_zL_c = 20\pi$  we observe the appearance of spatially distributed errors in the background, which we refer to as background noise.

The forcing modes are recovered directly from the SVD of the low-dimensional subspace projection. As they are used to recover the response modes using the linear operator, the accuracy of the forcing modes affect the results of the response ones. In the particular case of $St = 1$ and $k_zL_c = 20\pi$, when forcing mode is affected by noise, the randomized approach returns some structures emanating from the trailing edge in the response mode $\hat{\pmb{u}}_1^\text{rand}$, which was not present from the full resolvent analysis. This behavior is related to leakage from high-order modes, as $\sigma_1$ and $\sigma_2$ are close in the energy spectrum.
This remarkable level of agreement over all frequencies and spanwise wavenumbers ensures that the randomized approach presented in Algorithm 1 can help extract insights into the spatial structures to identify regions of sensitivity and guide flow control efforts.

\begin{figure}[t]
\footnotesize
 	\begin{tabular}{	 
						>{\centering\arraybackslash} m{0.30in}
						>{\centering\arraybackslash} m{2.01in}
						c
						>{\centering\arraybackslash} m{2.01in}
						c
						>{\centering\arraybackslash} m{2.01in}
					}
        	&{\bf  Full resolvent analysis}\vspace{-0.00in}
        	&
        	&{\bf  $\bm{U} = \bm{Q}\bm{\tilde{U}}$ (\citet{Halko:SIAMReview11})}\vspace{-0.00in}
        	&
        	&{\bf $\bm{U} = \bm{A} \bm{V} \bs{\Sigma}^{-1}$ (present)}\\
	        \cline{2-2}\cline{4-4}\cline{6-6}\\
             \vspace{-0.08in}$St$
			&\vspace{-0.08in}Response mode $\hat{\pmb{u}}_1^\text{full}$
			& 
			&\vspace{-0.08in}Response mode $\hat{\pmb{u}}_1^\text{rand}$
			&
			&\vspace{-0.08in}Response mode $\hat{\pmb{u}}_1^\text{rand}$\\
    \hline
                \vspace{+0.04in} $0.5$
			&	\vspace{+0.04in} \includegraphics[width=2.00in]{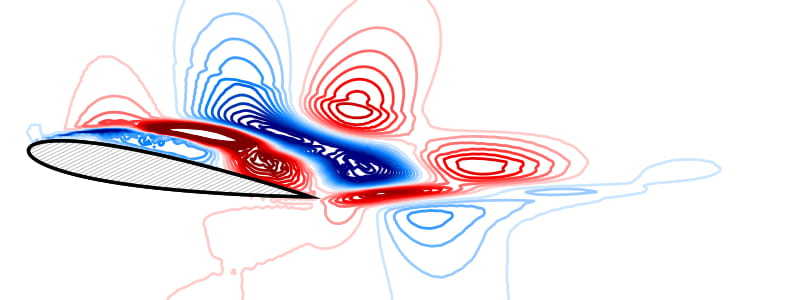}
			&
			&	\vspace{+0.04in} \includegraphics[width=2.00in]{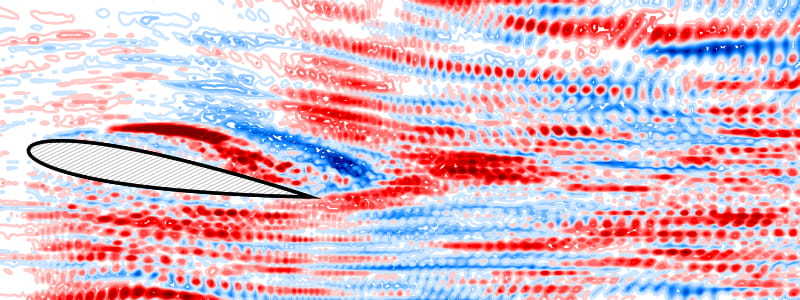}
			&
			&	\vspace{+0.04in} \includegraphics[width=2.00in]{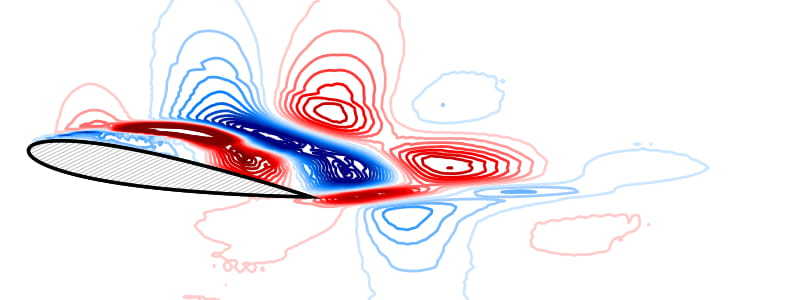}\\
    		    \vspace{+0.04in} $15$
    		&	\vspace{+0.04in} \includegraphics[width=2.00in]{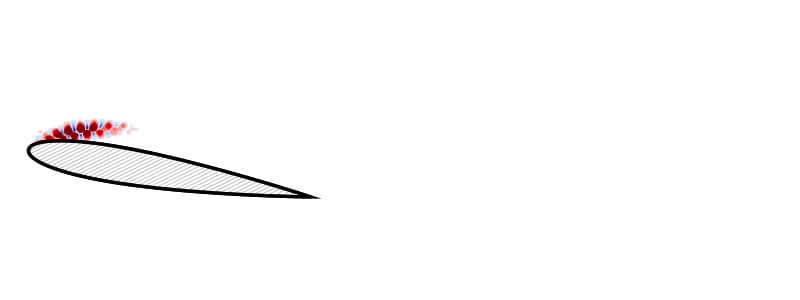}
			&
			&	\vspace{+0.04in} \includegraphics[width=2.00in]{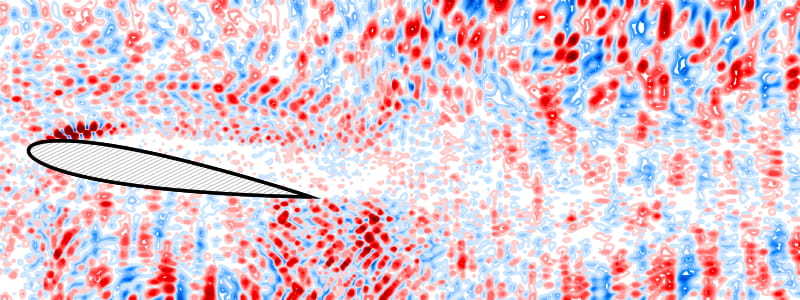}
			&
			&	\vspace{+0.04in} \includegraphics[width=2.00in]{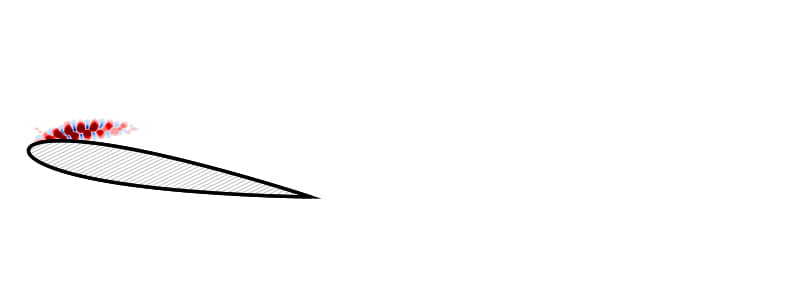}\\
	\end{tabular}
\caption{Comparison of the leading response modes ($\hat{\pmb{u}}_1$) recovered from the randomized resolvent analyses using $\bm{U} = \bm{Q}\bm{\tilde{U}}$ (the original approach in \citet{Halko:SIAMReview11}) and $\bm{U} = \bm{A} \bm{V} \bs{\Sigma}^{-1}$ (present, see equation (\ref{eq:U=AVS})).  The response modes from the full analysis are also shown for reference.  Results are shown for $k_zL_c = 0$ at representative frequencies $St$.  Modes are visualized with the streamwise velocity component with contour levels of $[-0.6, 0.6]$.} 
\label{fig:plot_mode_stand}
\end{figure}

These results were obtained using the present implementation that extends the original randomized SVD algorithm. In figure \ref{fig:plot_mode_stand}, using the original randomized algorithm \citep{Halko:SIAMReview11} within the resolvent analysis to recover the left singular vectors and singular values, the response modes contain background noise. Here, we use Algorithm 1 up to line 5, then SVD is performed as $\bm{B} = \bm{\tilde{U}}\bs{\Sigma}\bm{V}^*$ and $\bm{U}$ is recovered a posteriori using the original procedure for the randomized SVD with the resolvent analysis \citep{Halko:SIAMReview11,Moarref:JFM2013}, by $\bm{U} = \bm{Q}\bm{\tilde{U}}$. \citet{Yeh:JFM19} showed that the present problem setup presents a peak in the singular values near $St = 6$ for both spanwise wavenumbers, influenced by the eigenmodes associated with the shear-layer structure over the separation bubble. These eigenmodes are highly nonnormal and induce high-energy amplification through pseudoresonance \citep{Trefethen:Science93}. For the frequencies in a narrow region near $St = 6$, both implementations present similar results. Far from this band, the response modes obtained by the original procedure \citep{Halko:SIAMReview11} are contaminated by random background noise or leakage from higher-order modes, as shown in figure \ref{fig:plot_mode_stand} for $St = 0.5$ and $15$ and $k_zL_c = 0$. In these critical cases, the original randomized approach may not provide meaningful insights into flow physics. The present implementation shown in Algorithm 1 improves solving for the response modes.   Algorithm 1 does not enhance the forcing modes, as they are already accurate. When utilizing this technique to generate reduced-order models, one may perform steps 11 and 12 in Algorithm 1 to orthogonalize the left singular vectors. Considering the results obtained by our implementation, the modes are found very accurately for almost all frequencies and wavenumbers. To provide a concrete assessment, we quantitatively assess the accuracy of the randomized resolvent analysis. 

\begin{figure}[t]
\footnotesize
	\begin{overpic}[scale=0.20]{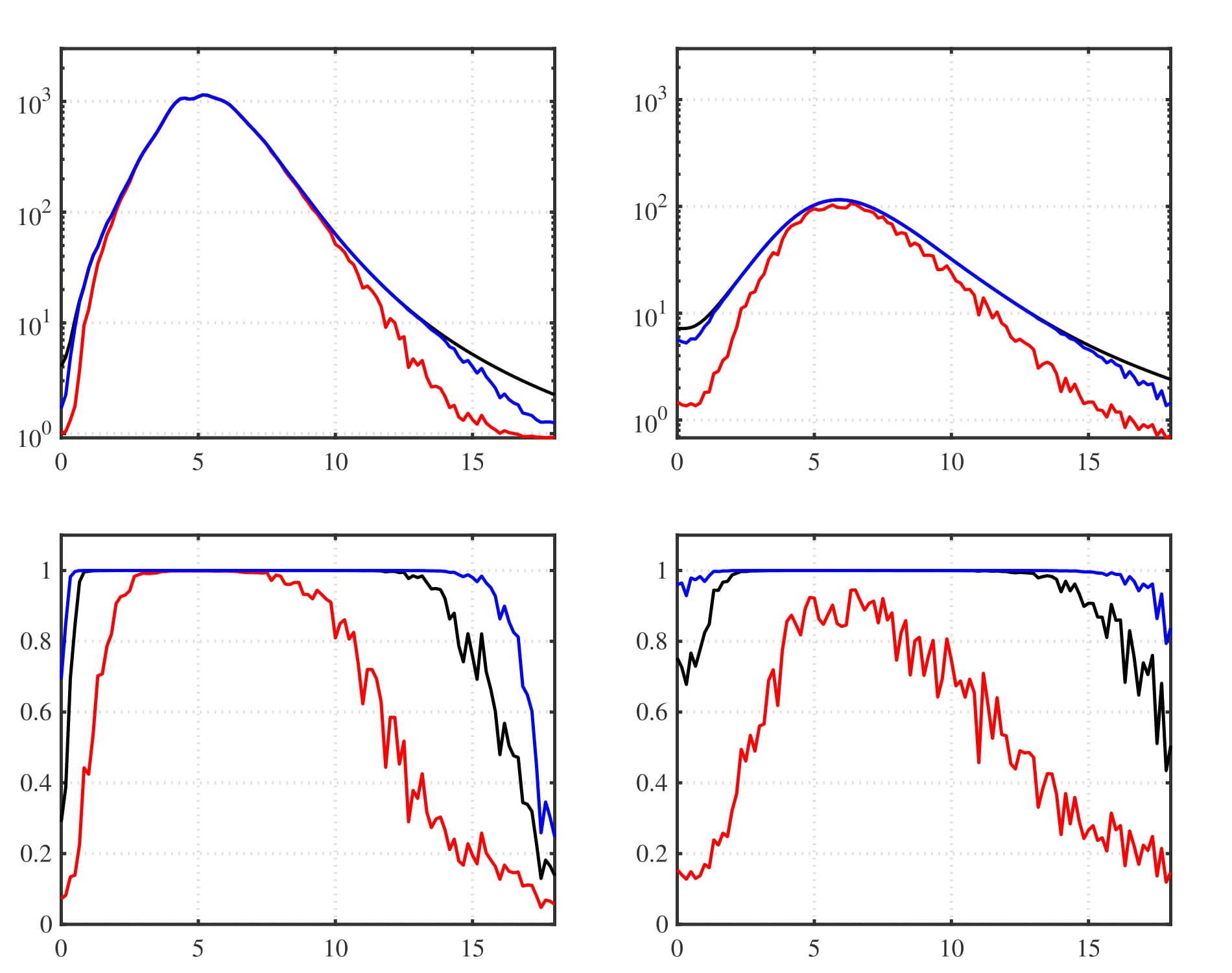}
		\put( 1, 76){(a)}
		\put(51, 76){(b)}
		\put( 1, 36.5){(c)}
		\put(51, 36.5){(d)}
		
		\put(20, 76.5){$k_zL_c = 0$}
		\put(70, 76.5){$k_zL_c = 20\pi$}
		
		\put(07, 48){
		    \begin{tabular}{l} 
			    {\color{black}Full analysis}\\
			    {\color{blue}Randomized (present)}\\
			    {\color{red}Randomized (original \citep{Halko:SIAMReview11})}\\
			\end{tabular}}

		\put(07, 09){
		    \begin{tabular}{l} 
			    {\color{black}Forcing}\\
			    {\color{blue}Response (present)}\\
			    {\color{red}Response (original \citep{Halko:SIAMReview11})}\\
		    \end{tabular}}

		\put(-3, 52){\rotatebox{90}{$\sigma_1^\text{full}$,~~$\sigma_1^\text{rand}$}}
		\put(-3, 6 ){\rotatebox{90}{$\langle \hat{\bs{u}}_1^\text{full}, \hat{\bs{u}}_1^\text{rand}\rangle$,~~$\langle \hat{\bs{v}}_1^\text{full}, \hat{\bs{v}}_1^\text{rand}\rangle$}}
		\put(16, -2){$St = \omega L_c/2\pi v_\infty$}
		\put(66, -2){$St = \omega L_c/2\pi v_\infty$}
	\end{overpic}
	\caption{(a,b) The leading amplification for the full 
	and randomized resolvent analyses. (c-d) Cosine similarities for the leading response $\langle \hat{\pmb{u}}_1^\text{full}, \hat{\pmb{u}}_1^\text{rand}\rangle$ and forcing $\langle \hat{\pmb{v}}_1^\text{full}, \hat{\pmb{v}}_1^\text{rand}\rangle$ modes.  Improvements of accuracy in the present randomized analysis can be observed by comparing the $\sigma_1^\text{rand}$ and $\hat{\bs{u}}_1^\text{rand}$ recovered from equation (\ref{eq:U=AVS}) to those from the original approach of \citet{Halko:SIAMReview11} (note that forcing modes obtained from both approaches are identical).  Results for spanwise wavenumber of $k_zL_c = 0$ and $20\pi$ are shown in the left and right columns, respectively.}
\label{fig:Gain_CosineSim}
\end{figure}

The agreement between the full and randomized analyses with respect to the gain (leading singular value) and modes over a range of frequencies is presented in figure \ref{fig:Gain_CosineSim}.  When Algorithm 1 is applied to recover left singular vectors and singular values, the randomized analysis accurately captures the trend of gain distribution over $1 \lesssim St \lesssim 15$ in figure \ref{fig:Gain_CosineSim}(a,b) for both wavenumbers. At the low and high-frequency ends, the gain shows deviations. The resemblance of the modal structure is quantified in figure \ref{fig:Gain_CosineSim}(c,d) with the cosine similarities, i.e., the inner products, $\langle \hat{\pmb{u}}_1^\text{full}, \hat{\pmb{u}}_1^\text{rand}\rangle$ and $\langle \hat{\pmb{v}}_1^\text{full}, \hat{\pmb{v}}_1^\text{rand}\rangle$.  As singular vectors are normalized, the cosine similarity of $1$ suggests that perfect match is attained between the modes from full and randomized resolvent analyses. Since these modes are complex, the cosine similarity removes dependence on the phase difference.
For almost the entire range of frequencies the cosine similarities are near unity, which means the agreement between full and randomized modes is excellent.  
When this value is reduced, the modes may be affected by noise, as seen for $\hat{\pmb{v}}_1^\text{rand}$ at $St = 1$ and $k_zL_c = 0$ in figure \ref{fig:plot_mode_stand}. For the original approach, using the randomized SVD algorithm \citep{Halko:SIAMReview11}, the modes have good agreement for a narrow band of frequencies only. By comparing the results from figures \ref{fig:plot_mode_avs} and \ref{fig:plot_mode_stand} to the values in figure \ref{fig:Gain_CosineSim}(c,d), we observe that the noise affects the modes when cosine similarity is below $0.5$. 
For frequencies and wavenumbers with cosine similarity up to 0.8 or higher, there is no noise and the results for full and randomized resolvent agree well. 
For this reason, it is desirable to search for solutions that provide a reliable agreement up to this scale to a broad range of frequencies and both wavenumbers. 
With the high-gain frequency range well captured, randomized resolvent analysis has demonstrated its capability of predicting the dominant pathway for energy amplification over the spectral space with reduced computational cost.

\begin{figure}
	\begin{overpic}[scale=0.20]{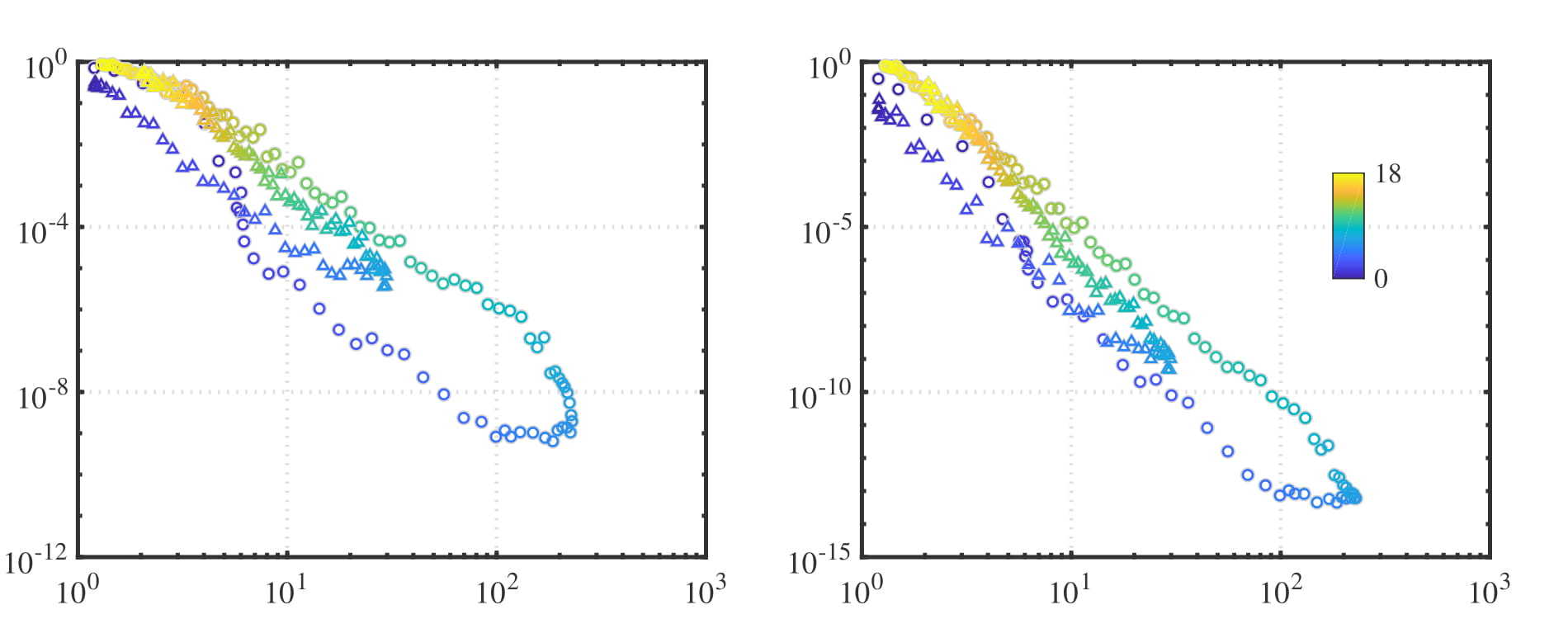}
    \footnotesize
		\put( 1, 37.5){(a)}
		\put(51, 37.5){(b)}
		\put(-3, 12){\rotatebox{90}{$1 - \langle \hat{\pmb{v}}_1^\text{full}, \hat{\pmb{v}}_1^\text{rand}\rangle$}}
		\put(47, 12){\rotatebox{90}{$1 - \langle \hat{\pmb{u}}_1^\text{full}, \hat{\pmb{u}}_1^\text{rand}\rangle$}}
		\put(19.5, -1.5){$(\sigma_1/\sigma_2)^\text{full}$}
		\put(69.5, -1.5){$(\sigma_1/\sigma_2)^\text{full}$}
		\put(84.75, 29.5){$St$}
	\end{overpic}
	\caption{Error based on cosine similarity for the leading (a) forcing modes ($1 - \langle \hat{\pmb{v}}_1^\text{full}, \hat{\pmb{v}}_1^\text{rand}\rangle$) and (b) response modes ($1 - \langle \hat{\pmb{u}}_1^\text{full}, \hat{\pmb{u}}_1^\text{rand}\rangle$) over the leading gap from the full resolvent analysis. 
	The symbols $\circ$ and {\scriptsize $\triangle$} represent results from $k_zL_c = 0$ and $20\pi$, respectively, colored by $St$.}
    \label{fig:Gain_Gap}
\end{figure}

As stated in section \ref{sec:random}, the use of low-rank approximation in the randomized approach is built upon the assumption of the low-rank nature of the resolvent operator.  
The randomized resolvent analysis shows its strength when the singular values exhibit fast decay, as evident from figure \ref{fig:Gain_Gap}.  The accuracy of the modal structured captured by randomized analysis is examined with respect to the ratio of the leading and second singular values, $(\sigma_1/\sigma_2)^\text{full}$ from the full resolvent analysis.  
The error in the modal structures exhibits a decreasing trend as this ratio increases. 
When this ratio is close to unity, the randomized technique may not accurately separate the first and second modes.  In fact, the aforementioned trailing edge structure that appeared in the randomized response mode for $k_zL_c = 20\pi$ and $St = 1$ is caused by the leakage of the structures from the second response mode (see figure \ref{fig:plot_mode_avs}).  When the ratio $(\sigma_1/\sigma_2)^\text{full}$ is above $30$, the error decreases to $\lesssim 10^{-5}$ for forcing modes and $\lesssim 10^{-8}$ for response modes. 

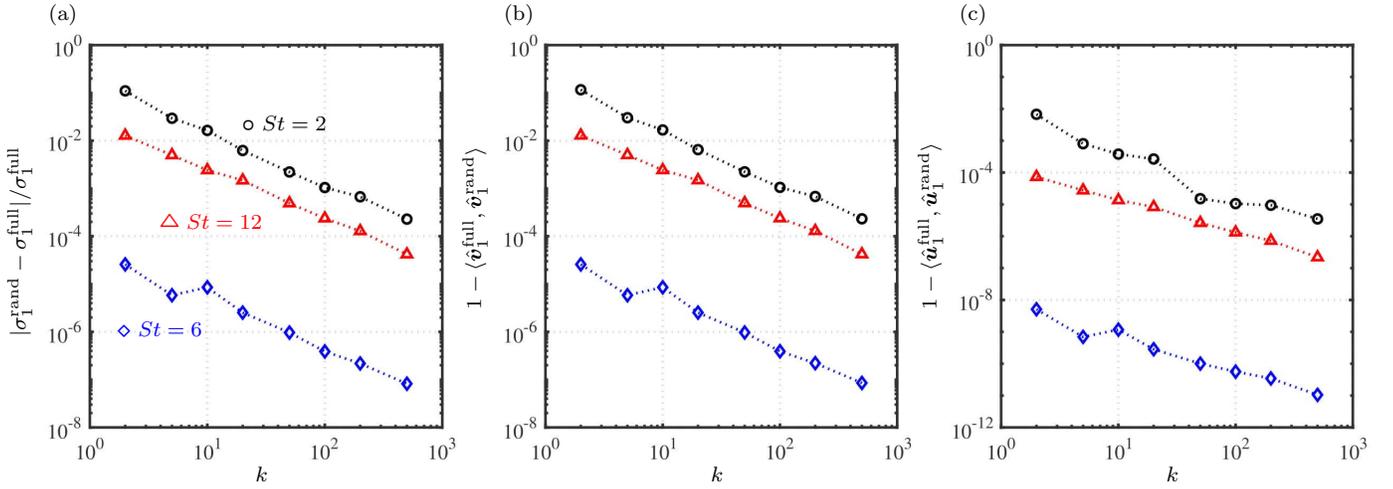
\begin{figure}[t]
	\begin{overpic}[scale=0.20]{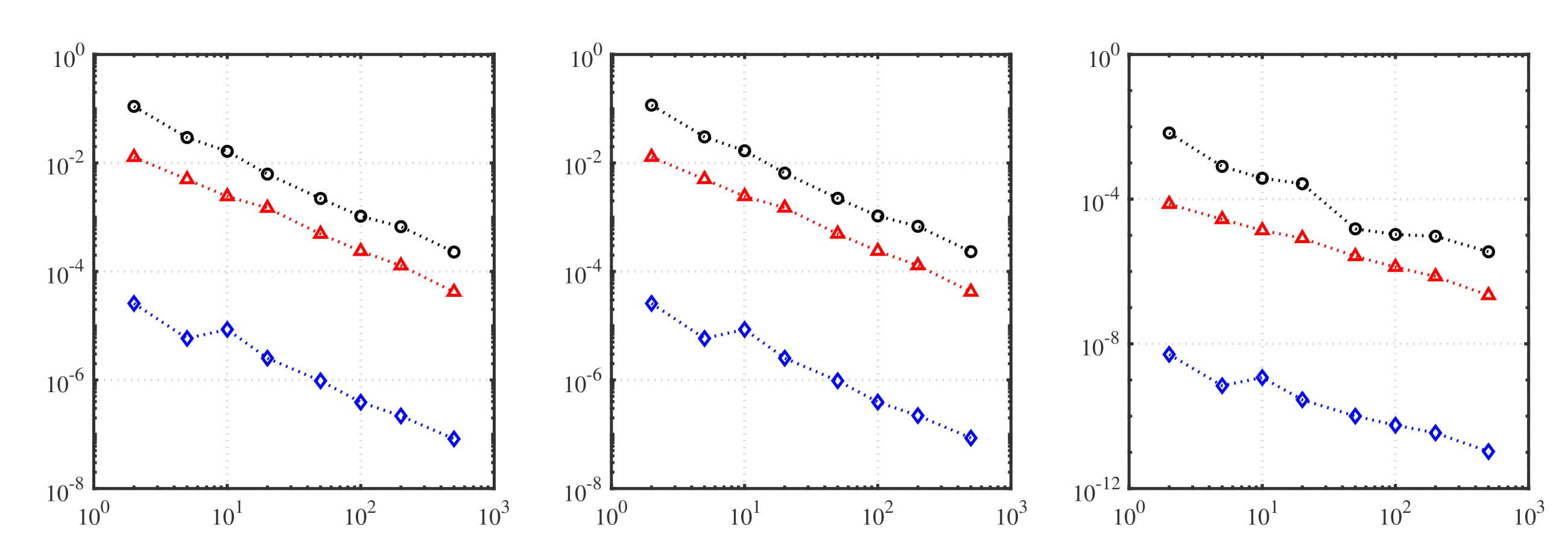}
	\footnotesize
		\put(3, 33){(a)}
		\put(36, 33){(b)}
		\put(69, 33){(c)}
		\put( 0, 10){\rotatebox{90}{$|\sigma_1^{\text{rand}} - \sigma_1^{\text{full}}| / \sigma_1^{\text{full}}$}}
		\put( 33, 12){\rotatebox{90}{$1 - \langle \hat{\bs{v}}_1^\text{full}, \hat{\bs{v}}_1^\text{rand}\rangle$}}
		\put( 66, 12){\rotatebox{90}{$1 - \langle \hat{\bs{u}}_1^\text{full}, \hat{\bs{u}}_1^\text{rand}\rangle$}}
		\put(18, -0.5){$k$}
		\put(51, -0.5){$k$}
		\put(84, -0.5){$k$}
		\put(17, 25){$\boldsymbol{\circ}$ $St = 2$}
		\put(8, 10){${\color{blue}\boldsymbol{\diamond}}$ \color{blue}$St = 6$}
		\put(11, 18){\color{red}\scriptsize $\boldsymbol{\triangle}$} 
		\put(13, 17.85){\color{red} $St = 12$}
	\end{overpic}
	\caption{Influence of test matrix size $k$ on the accuracy of (a) leading singular value, (b) forcing mode, and (c) response mode at $k_zL_c = 20\pi$.  All exhibit $\mathcal{O}(k)$ convergence.}
    \label{fig:plot_kinfluence}
\end{figure}

Next, we study the influence of the width of the test matrix $k$ on the error in the leading singular values and modes, as presented in figure \ref{fig:plot_kinfluence}.  When the value of $k$ is varied from $2$ to $500$, the error from the use of randomized analysis decreases.  For three representative frequencies, we observe the same rate of convergence $\approx \mathcal{O}(k)$ for both the gain and cosine similarity.  As stated in section \ref{sec:random}, increasing $k$ has the same practical effect of oversampling, in the present application.
For this flow, we observe that $k = 10$ is sufficient to achieve sufficient accuracy with $\lesssim 1 \%$ error, which is remarkably low when compared to the high dimensionality of the resolvent operator.

The computational cost of the randomized resolvent technique can be further reduced. For instance, the biconjugate gradient stabilized (BiCGStab) or the generalized minimum residual (GMRES) methods can be utilized to solve linear systems with appropriate preconditioners (e.g., incomplete LU and Jacobian). 


\subsection{Higher-order modes}
\label{sec:high}

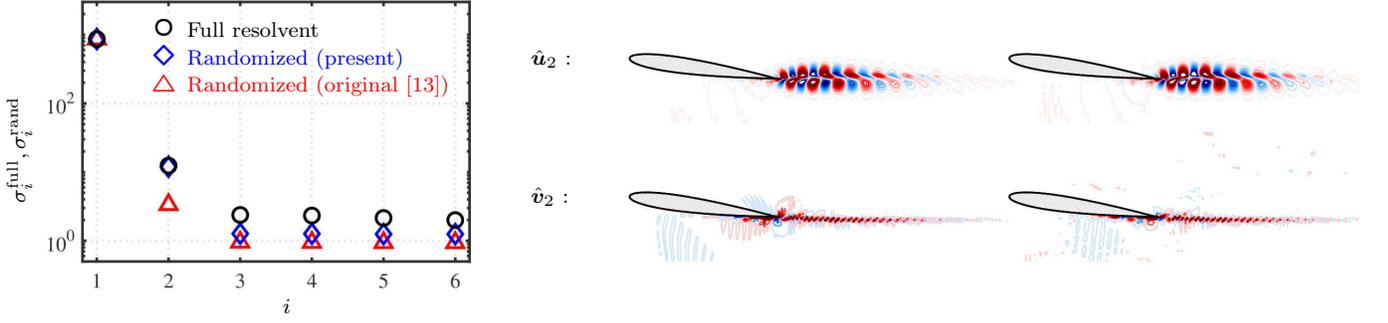
\begin{figure}[t]
\footnotesize
    \vspace{0.0in}
    \begin{overpic}[scale=0.20]{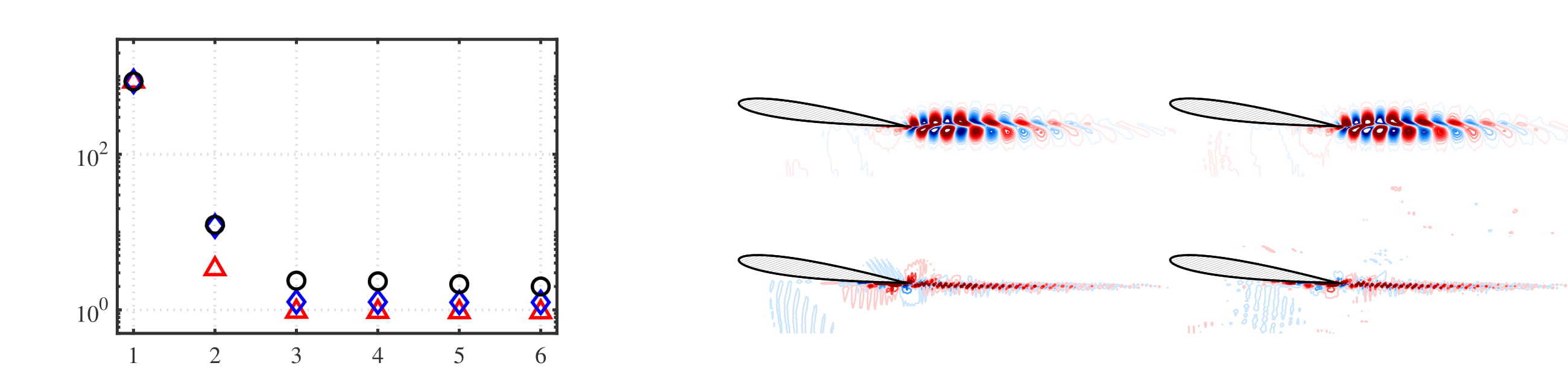}
		\put(50, 23.5){\bf Full Resolvent}
		\put(75, 23.5){\bf Randomized (present)}
		\put( 2, 8){\rotatebox{90}{$\sigma_i^{\text{full}},\sigma_i^{\text{rand}}$}}
		\put( 22, 0){$i$}
		\put( 40,  8){$\hat{\bs{v}}_2$ :}
		\put( 40, 18){$\hat{\bs{u}}_2$ :}
	\small
        \put(11.5, 18){
            \begin{tabular}{c l}
            	{\color{black}\Large    $\boldsymbol{\circ}$}       & {\color{black}\footnotesize Full resolvent} \\ 
				{\color{blue}\Large     $\boldsymbol{\diamond}$}    & {\color{blue}\footnotesize Randomized (present)}\\
				{\color{red}           $\boldsymbol{\triangle}$}   & {\color{red}\footnotesize Randomized (original \citep{Halko:SIAMReview11})} \\ 
			\end{tabular}
        }
	\end{overpic}
	\caption{Recovery of the higher-order resolvent gains ($\sigma_{i = 1-6}^{\text{rand}}$) and modes ($\hat{\bs{v}}_2$ and $\hat{\bs{u}}_2$) for $St = 4$ and $k_zL_c = 0$.  The resolvent gains obtained from the present randomized analysis ({\color{blue}\large$\boldsymbol{\diamond}$}, recovered by equation (\ref{eq:U=AVS})) and those from the original approach of \citet{Halko:SIAMReview11} ({\color{red}$\boldsymbol{\triangle}$}) are compared to those from the full resolvent analysis ({\color{black}\large$\boldsymbol{\circ}$}). }
    \label{fig:highorder}
\end{figure}

Let us discuss the performance of the randomized technique with respect to the high-order modes. For some cases, the second largest singular value may also be spaced apart from the higher-order singular values and be determined accurately. In figure \ref{fig:highorder}, we show for $k_zL_c = 0$ and $St = 4$ a case where both the leading and second singular values are spread from the rest of the singular values. In this case, the randomized algorithms accurately capture the second modes. The flow structures at the trailing edge are perceived in the response modes. The forcing modes appear over the pressure side near the trailing edge. For the results obtained from Algorithm 1, the modes are the same for randomized resolvent and for the full resolvent. However, for the resolvent analysis using the randomized SVD algorithm \citep{Halko:SIAMReview11} within the resolvent analysis, the secondary singular values are poorly captured and the modes are polluted by background noise and leakage from other modes (not shown here). For this reason, when applying randomized resolvent, the present implementation shown in Algorithm 1 must be considered as they can approximate the detached high-order modes accurately.

\subsection{Choice of the test matrix $\bs{\Omega}$}
\label{sec:physicsomega}

For the randomized resolvent analysis with a test matrix size of $k = 10$, Gaussian, orthonormal, Rademacher and ultrasparse Rademacher test matrices provide similar results and no observable difference in computational savings. By using the implementation shown in Algorithm 1, all test matrices present similar accuracy as shown in figures \ref{fig:Gain_CosineSim} and \ref{fig:Gain_Gap}. For very low and very high $St$ numbers in the range of frequencies analyzed in this work, where the singular values decay slowly, one can increase the size $k$ or apply subspace or power iterations.  However, it is possible to obtain more accurate results by constructing a random test matrix $\bs{\Omega}$ that incorporates physical insights from the base flow.

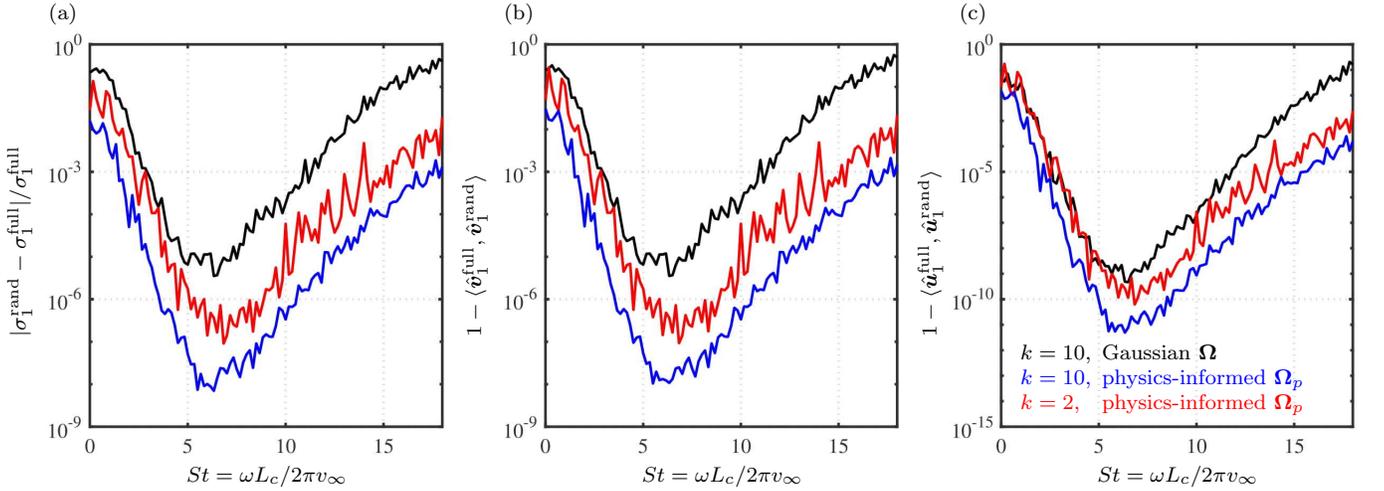
\begin{figure}
	\begin{overpic}[scale=0.20]{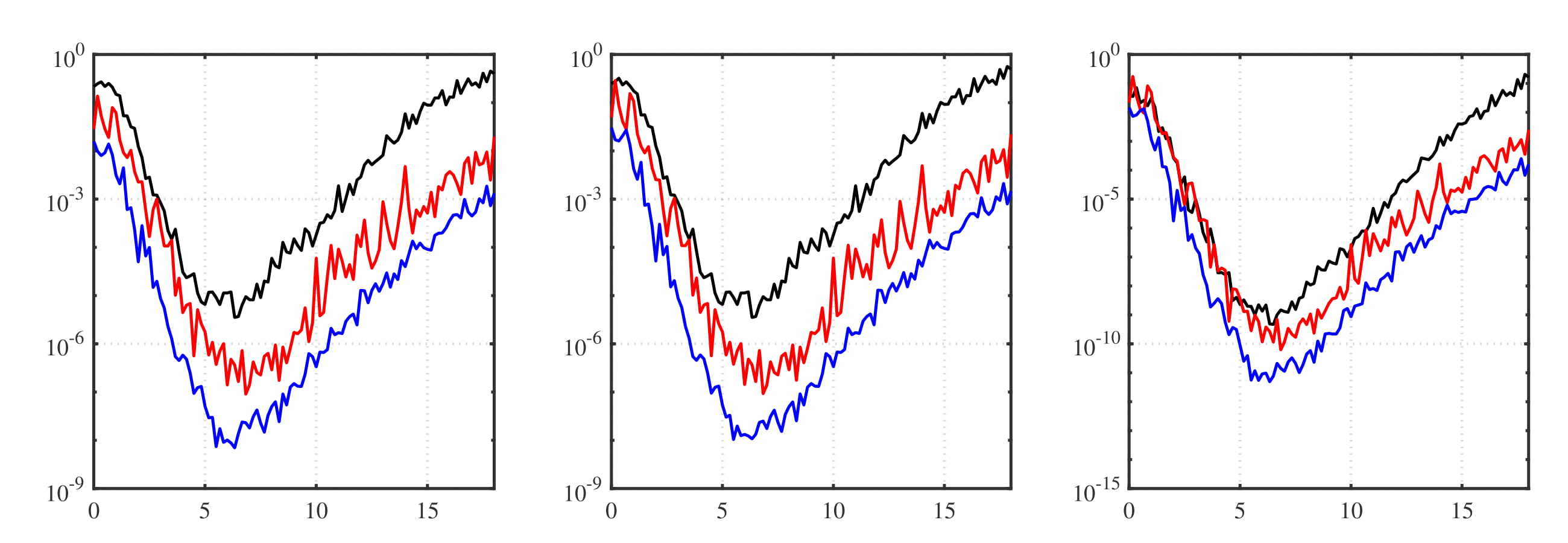}
    \footnotesize
		\put(3, 33){(a)}
		\put(36, 33){(b)}
		\put(69, 33){(c)}
		\put( 0, 10){\rotatebox{90}{$|\sigma_1^{\text{rand}} - \sigma_1^{\text{full}}| / \sigma_1^{\text{full}}$}}
		\put( 33, 10){\rotatebox{90}{$1 - \langle \hat{\bs{v}}_1^\text{full}, \hat{\bs{v}}_1^\text{rand}\rangle$}}
		\put( 66, 10){\rotatebox{90}{$1 - \langle \hat{\bs{u}}_1^\text{full}, \hat{\bs{u}}_1^\text{rand}\rangle$}}
		\put(13, -0.5){$St = \omega L_c/2\pi v_\infty$}
		\put(46, -0.5){$St = \omega L_c/2\pi v_\infty$}
		\put(79, -0.5){$St = \omega L_c/2\pi v_\infty$}
		\put(72.5, 6.5){
		    \begin{tabular}{l l}
                {\color{black}$k=10$,} & {\color{black}Gaussian $\bs{\Omega}$} \\ 
				{\color{blue}$k=10$,}    & {\color{blue}physics-informed $\bs{\Omega}_p$}\\
				{\color{red}$k=2$,}      & {\color{red}physics-informed $\bs{\Omega}_p$} \\ 
			\end{tabular}
		}
		
	\end{overpic}
	\caption{Influence of test matrices on the relative error for (a) gain distribution and cosine similarities for (b) forcing and (c) response modes, $\hat{\pmb{v}}_1$ and $\hat{\pmb{u}}_1$, at $k_zL_c = 20\pi$.  Results are shown for Gaussian random normal distribution test matrix {\color{black}$\bs{\Omega}$} and physics-informed random test matrices {$\bs{\Omega}_p = {\sf{diag}}(\bs{\Phi})\bs{\Omega}$} with sizes $\color{blue}k = 10$ and $\color{red}k = 2$.
	}
    \label{fig:Gain_CosineSim_tm}
\end{figure}

While the random test matrix is effective in yielding accurate results, we can consider constructing a test matrix that can generate the entries in a smart manner by incorporating the knowledge of the base flow. We know that regions of strong shear are important in amplifying forcing inputs.  Moreover, regions with minimal velocity gradients are not that important. For these reasons, the velocity gradient at each grid point can be used to scale the test matrix.  Here, we propose a physics-informed test matrix, $\bs{\Omega}_p$, scaled by the 2-norm of the velocity gradient, $||\bs{\nabla} {\bm{v}}||_2$, where $ {\bm{v}}$ is the velocity vector. We construct a scaling factor $\bs{\Phi}_j = ||\bs{\nabla} {\bm{v}}_j||_2$ at each grid point $j$. The scaling vector $\bs{\Phi}$ has to be stacked according to the number of variables to reach the size $m$ of the linear operator.  The physics-informed test matrix then becomes
\begin{equation}
    \bs{\Omega}_p = {\sf{diag}}(\bs{\Phi})\bs{\Omega}.
    \label{eq:physics}
\end{equation}
The results based on the physics-informed test matrix are shown in figure \ref{fig:Gain_CosineSim_tm}.  While the results obtained from the use of a normal distribution test matrix have shown excellent accuracy, the use of physics-informed test matrix further improves the accuracy by a few orders of magnitude for the considered frequencies and spanwise wavenumbers.  More importantly, we achieve results with comparable or higher level of accuracy using a extremely low width of the test matrix of $k = 2$. This results in a considerable reduction in computation time (see Table \ref{tab:comp}). Using $k = 10$, linear systems are solved at least 30 times. Now, using $k = 2$, only 6 linear solvers are needed to obtain the same accuracy, which is achieved only with a physics-informed scaling of the test matrix.  By combining randomized numerical linear algebra and some physical insights, we are now empowered to perform the input-output analysis for ever more complex 2D and 3D turbulent base flows on a standard computer, or on a high-performance computing cluster to expand the envelope of resolvent analysis.


\section{Conclusion}
\label{sec:conclusion}

Resolvent analysis has proven to be a powerful technique to reveal the input-output characteristics of fluid flows.  However, the computational cost and memory allocation of the resolvent analysis can be taxing for high-Reynolds number flows, making it prohibitive to be applied to complex turbulent base flows.  The major computational cost of the analysis is associated with the SVD of the resolvent operator.  To remove this bottleneck, the randomized approach has been adopted to reduce the computational cost of SVD by considering the low-rank approximation of the resolvent operator.
This was achieved by constructing the low-rank basis based on the insights gained from the sketch of the resolvent, which is obtained from a linear system solver.
For flows with fast singular value decays, e.g., flows with strong shear and separation, the randomized resolvent analysis reveals its power to accurately capture the response and forcing modes as well as the gain.  
Moreover, we consider the use of the velocity gradient to scale the random test matrix. Such scaling enhanced the accuracy of the randomized resolvent analysis. The necessary computation time was significantly reduced as the number of linear systems to be solved is considerably smaller.  To demonstrate the capability of the randomized resolvent analysis, we analyzed the turbulent separated flow over a NACA 0012 airfoil.  Excellent agreement of the leading forcing and response modes and the gains were shown between the full and randomized resolvent methods.  By incorporating the knowledge of the base flow in terms of its velocity gradient into the randomized test matrix, additional speed up and accuracy enhancement were achieved.  
With the computational cost and memory allocation being relieved with the randomized approach, the application of the resolvent analysis can be significantly extended to higher-Reynolds number 2D and 3D base flows.


\section*{Acknowledgments}

We gratefully acknowledge the support from the Office of Naval Research (N00014-16-1-2443), Air Force Office of Scientific Research (FA9550-18-1-0040), and Army Research Office (W911NF-17-1-0118).  We thank L. Mathelin, S. L. Brunton, and N. B. Erichson for the enlightening discussions.  Some of the computations presented here were supported by the Department of Defense High Performance Computing Modernization Program.  



\end{document}